%% file: p.tex
\documentclass[conference]{sty/IEEEtran}
\IEEEoverridecommandlockouts

\include{pkgs}

\def\BibTeX{{\rm B\kern-.05em{\sc i\kern-.025em b}\kern-.08em
    T\kern-.1667em\lower.7ex\hbox{E}\kern-.125emX}}

\newcommand{\sys}{\mbox{\textsc{SoK}}\xspace}

\newcount\countAAA
\newcounter{xkcount}[section]

\newcounter{todocount}

\input{cmds}
\input{rev}

\begin{document}

\input{hdr}
\date{}
\maketitle

\input{abstract}
\input{intro}
\input{architecture}

\input{sok}
%         % \input{casestudy}
\input{sok-2}

\input{pwn2own}
\input{discussion}

% %         %\input{proposal}

% \input{lessons}
\input{conclusion}

% \begin{enumerate}
%     \renewcommand*{\do}[1]
%     {\advance\countAAA by 1
%     \item \textcolor{blue}{[TODO-\the\countAAA: #1]}}
%     \dolistloop{\todolists}
% \end{enumerate}

% \setlength{\bibsep}{0.0pt}
\bibliographystyle{plain}
\scriptsize
\bibliography{p,sslab,conf,mitigations}

\input{appendix}

\end{document}

%% file: pkgs.tex
\input{warning}

\usepackage[hyphens]{url}
\usepackage[breaklinks,colorlinks]{hyperref}
\hypersetup{citecolor=blue,linkcolor=blue}
\usepackage{tabularx}

\usepackage{threeparttable}

\usepackage{cite}
\usepackage{amsmath,amsopn,amssymb}
\usepackage{endnotes,microtype,xspace,graphicx,fancyvrb,multirow}
\usepackage{booktabs}
\usepackage{array,underscore,relsize}
\usepackage[T1]{fontenc}
\usepackage{times}
\usepackage{fancyhdr,lastpage}
\usepackage{enumitem}
\usepackage[labelfont=bf,font=small,skip=5pt]{caption}
\usepackage{listings}
\usepackage{rotating}
\usepackage{pifont}
\usepackage{comment}
\usepackage{colortbl}
\usepackage{adjustbox}
\usepackage{subcaption}

\usepackage{fp}
\usepackage{siunitx}
\usepackage{mdframed}

\usepackage{epstopdf}

\usepackage{balance}

%% file: warning.tex
\usepackage{silence}

%% file: cmds.tex
\newcommand{\cc}[1]{\mbox{\smaller[0.5]\texttt{#1}}}

\fvset{fontsize=\scriptsize,xleftmargin=8pt,numbers=left,numbersep=5pt}

\def\Snospace~{\S{}}

\newcommand{\G}{\cellcolor[rgb]{0.85,0.85,0.85}}

\newcommand{\red}[1]{\textcolor{red}{#1}}
\newcommand{\blue}[1]{\textcolor{blue}{#1}}
\newcommand{\green}[1]{\textcolor[rgb]{0.1,0.54,0.1}{#1}}

\newcommand{\yes}{\ding{51}}
\newcommand{\no}{\ding{55}}
\newcommand{\start}{\ding{72}}

\newcommand{\x}{$\times$\xspace}

\if 0

\fi

\newif\ifdraft\drafttrue
\newif\ifnotes\notestrue
\ifdraft\else\notesfalse\fi

\input{glyphtounicode}
\newcolumntype{R}[1]{>{\raggedleft\let\newline\\\arraybackslash\hspace{0pt}}p{#1}}

\newcommand{\includepdf}[1]{
  \includegraphics[width=\columnwidth]{#1}
}

\newcommand{\squishlist}{
\begin{itemize}[noitemsep,nolistsep]
  \setlength{\itemsep}{-0pt}
}
\newcommand{\squishend}{
  \end{itemize}
}

\lstdefinelanguage{JavaScript}{
  morekeywords=[1]{break, continue, delete, else, for, function, if, in,
    new, return, this, typeof, var, void, while, with},
  morekeywords=[2]{false, null, true, boolean, number, undefined,
    Array, Boolean, Date, Math, Number, String, Object, mov, eax, ax, bx},
  morekeywords=[3]{eval, parseInt, parseFloat, escape, unescape},
  sensitive,
  morecomment=[s]{/*}{*/},
  morecomment=[l]//,
  morecomment=[s]{/**}{*/}, % JavaDoc style comments
  morestring=[b]',
  morestring=[b]"
}[keywords, comments, strings]

\makeatletter
\newcommand\BeraMonottfamily{%
  \def\fvm@Scale{0.85}% scales the font down
  \fontfamily{fvm}\selectfont% selects the Bera Mono font
}
\makeatother
\lstdefinestyle{JavaScriptStyle}{
language=JavaScript,
frame=single,
basicstyle=\BeraMonottfamily\scriptsize,
keywordstyle=\color{blue},
literate = {-}{-}1,
aboveskip=3pt,belowskip=2pt
}

\usepackage{tikz}

\usepackage{xstring}
\newcommand{\PP}[1]{
\vspace{2px}
\noindent{\bf \IfEndWith{#1}{.}{#1}{#1.}}
}

\newcommand{\PPi}[1]{
\vspace{2px}
\noindent{\textit{\IfEndWith{#1}{.}{#1}{#1.}}}
}

\newcommand{\ff}{Firefox\xspace}
\newcommand{\safari}{Safari\xspace}
\newcommand{\chrome}{Chrome\xspace}
\newcommand{\edge}{Edge\xspace}
\newcommand{\js}{JavaScript\xspace}

\newcommand{\macos}{macOS\xspace}

\newcommand{\ie}{\textit{i}.\textit{e}.}
\newcommand{\eg}{\textit{e}.\textit{g}.}

\newcounter{lessoncount}
\newcommand{\lesson}[2]{
\refstepcounter{lessoncount}
\vspace{4pt}
\noindent\fbox{\parbox{0.96\linewidth}{
    \textbf{Lesson~\thelessoncount: \IfEndWith{#1}{.}{#1}{#1.}}
\newline{#2}
}}}

\newcommand{\boxbeg}{
\noindent
\vspace{2px}
\begin{tabular}{|l|}\hline
\begin{minipage}{0.94\columnwidth}
\vspace{2px}
\noindent
}

\newcommand{\boxend}{
\vspace{2px}
\end{minipage}\\ \hline
\end{tabular}
\vspace{2px}
}

%% file: rev.tex
\gdef\therev{732e8f5}
\gdef\thedate{2021-12-31 11:44:16 -0500}

%% file: hdr.tex
\title{\sys: On the Analysis of Web Browser Security}

% when 'make draft'
\ifdefined\DRAFT
 \pagestyle{fancyplain}
 \lhead{Rev.~\therev}
 \rhead{\thedate}
 \cfoot{\thepage\ of \pageref{LastPage}}
\fi

\author{
{\rm Jungwon Lim*,}\;
{\rm Yonghwi Jin*$^\dagger$,}\;
{\rm Mansour Alharthi,}\;
{\rm Xiaokuan Zhang,}\;
\\
{\rm Jinho Jung,}\;
{\rm Rajat Gupta,}\;
{\rm Kuilin Li,}\;
{\rm Daehee Jang$^\ddagger$,}\;
{\rm Taesoo Kim}\;
\\
{ Georgia Institute of Technology} \qquad{$^\dagger$Theori Inc.} \qquad{$^\ddagger$Sungshin Women's University}
}

%% file: abstract.tex
\begin{abstract}
Web browsers are integral parts of everyone's daily life.
They are commonly used
for security-critical and privacy sensitive tasks,
like banking transactions and checking medical records.
Unfortunately,
modern web browsers are
too complex to be bug free
(\eg, 25 million lines of code in Chrome),
and their role as an interface to the cyberspace
makes them an attractive target for attacks.
Accordingly,
web browsers naturally
become an arena for demonstrating
advanced exploitation techniques by attackers
and state-of-the-art defenses by browser vendors.
Web browsers,
arguably,
are the most exciting place to learn
the latest security issues and techniques,
but remain as a black art to
most security researchers
because of their fast-changing characteristics
and complex code bases.

To bridge this gap,
this paper attempts to systematize
the security landscape of modern web browsers
by studying the popular classes of security bugs,
their exploitation techniques,
and deployed defenses.
More specifically,
we first introduce
a unified architecture
that faithfully represents
the security design of four major web browsers.
Second,
we share insights
from a 10-year longitudinal study on browser bugs.
Third,
we present a timeline and context of mitigation schemes
and their effectiveness.
Fourth,
we share our lessons from
a full-chain exploit used in 2020 Pwn2Own competition.
We believe that
the key takeaways from this systematization
can shed light on
how to advance the status quo of modern web browsers,
and, importantly, how to create secure yet complex software
in the future.

\end{abstract}

%% file: intro.tex
\section{Introduction}
\label{s:intro}

Web browsers play an integral role
in the modern, Internet-connected lifestyle.
We rely on web browsers
to pay mortgages,
schedule vaccines,
and connect to people worldwide.
In other words,
web browsers become the gatekeeper
to cyberspace,
and their insecurity
is a critical threat to
the safety, fairness and privacy
of our society.
Unfortunately,
web browsers
have been the most attractive, valuable target
of cyber attacks---%
50\% of 0-day exploits found in the wild
were attacking web browsers
in 2021~\cite{p0-itw-bugs}
and threatened
\emph{every single} person on the Internet%
~\cite{ff-itw-tor,chrome-itw-owo, itw-series-chrome,
  chrome-itw-pzmk,msft-itw-mshtml,
  ios-itw-exploit,ios-itw-million-dollar}.

Accordingly,
modern web browsers naturally become a battlefield
for attackers
 who wish to break in with novel exploit techniques,
and browser vendors
 who want to keep users safe with
 the most advanced mitigation schemes.
Browser vendors
are indeed the essential players
that advance modern security practices by
1) open sourcing not only the current architecture and code
but also the design process itself~\cite{chr-src,chromium-home,ff-src};
2) introducing bug bounty awards
to encourage the discovery of 0-day bugs~\cite{chrome-bounty-program,ff-bounty-program};
and 3) proactively finding exploitable bugs
by developing and running state-of-the-art fuzzers
on the cloud~\cite{clusterfuzz,clusterfuzz-specs}.

Unfortunately,
detailed design decisions for security
and insights on new mitigations against novel exploitation
are often considered as a black art,
keeping their lessons learned within the web browser community.
This is partly because of
their complex architecture,
fast-changing implementation,
and overwhelming size of code bases,
but mainly because
it is non-trivial to systematize the knowledge
of all major web browsers
coherently and objectively
simultaneously.
Experts from each browser vendor have
attempted to provide their perspectives
on security design and decisions,
\eg, Chrome~\cite{chr-sec},
IE~\cite{edge-sec}, Firefox~\cite{ff-sec}.
Previous industry reports published in 
2017~\cite{browser-white-paper2,browser-white-paper1}
mainly focus on describing individual techniques 
and defenses in an ad-hoc, as-it-is manner
without developing academic perspectives
or providing insights and lessons
that are useful to envision the next directions for the community. 

This paper makes
a bold attempt to systematize the security landscape
of modern web browsers.
We first provide a unified model of four major web browsers
as they pertain to security,
and compare and contrast their security decisions
by using the provided model.
Second, based on the model,
we analyze security bugs found in each open source browser 
in the last 10 years,
and show their relation to
the development of new mitigation schemes,
bug bounty programs,
and known exploitation techniques used in the wild.
Third,
based on our study,
we convey our insights and lessons
to inspire researchers and developers
who are shaping the future of web browsers.
We hope that
our systematization attempt
can help them
to understand the approaches of each vendor in a holistic manner
and thus enhance their security designs
to minimize security impacts and attack surfaces.

% put it on the 2nd page 
\begin{figure*}[t!]
  \centering \includegraphics[width=0.95\textwidth]{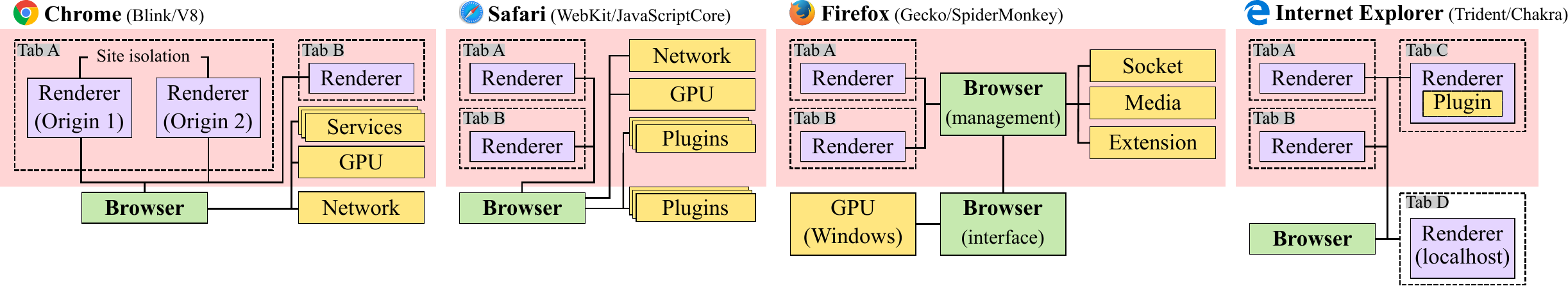}
  \caption{
    Internal architecture of four major web browsers.
    All browsers deploy a sandbox (pink region)
    to restrict the renderer, while the
    detailed sandboxing methodology
    differs based on the underlying OS.
    There are subtle but important
    differences across browsers.
    }
  \label{fig:arch}
  \vspace{-13pt}
\end{figure*}

\PP{Challenges}
Three unique characteristics 
make it
challenging to
systematize the knowledge of web browser security.

\begin{enumerate}[noitemsep,nolistsep]
  \setlength{\itemsep}{-0pt}
\item \textbf{A moving target.}
  Browser vendors make decisions rapidly (\eg, weekly updates)
  and their development is at a much faster pace
  than any other software that humans have built.
  To infer insightful lessons,
  we strive hard to stay focused on
  fundamental design issues and approaches
  in web browsers.
\item \textbf{Overwhelming size.}
  Modern web browsers are built with a few million lines of code,
  \eg, Chrome consists of 25 million lines of code~\cite{browser-LoC}.
  In addition to the project size,
  information on web browsers,
  such as 0-day exploits and mitigations,
  is scattered all over the Internet
  and fails to provide a holistic summary 
  and overview of the security landscape.
  In this paper,
  we limit our interest to the four major web browsers, namely, 
  \chrome, \ff, \safari, and \edge,
  and study multiple, public sources for their security issues:
  issue trackers~\cite{chr-bug-trac,ff-bug-trac},
  CVE reports~\cite{nvd,CVE-2016-4622,CVE-2013-6632,
  CVE-2006-5579,CVE-2017-5124,
  CVE-2019-6481,CVE-2020-6554,
  CVE-2021-34506,CVE-2020-6809},
  code repositories~\cite{ch,chr-src,v8-src,ff-src},
  and technical reports from the
  vendors~\cite{p0-bug-trac,p0-itw-bugs,
  webkit-css-jit,webkit-gpu-sb,ff-timer-spectre,chrome-timer-spectre,
  webkit-structureid,WDAG,miracleptr3,
  npapi-expkits-ms,android-bitmap-exploit,firefox-flash-sandboxing,
  itw-series-chrome,itw-tag-summary,ios-itw-exploit}.
\item \textbf{Unique designs.}
  It is also important to provide
  an objective perspective on their security issues; 
  each browser has its own restrictions and requirements
  in making decisions (\eg, release deadline),
  and it is critical to focus on the fundamental issues
  that our community can learn from.
  To solve this,
  we provide a unified architecture
  that compares and contrasts each browser's design
  conceptually
  without compromising their implementation details.
\end{enumerate}

\PP{Contributions}
This paper makes the following contributions:
\squishlist
  \item Provide a detailed comparison of modern browser architectures
    in terms of sandboxing schemes;
  \item Present a 10-year longitudinal study on browser bugs;
  \item Categorize browser vulnerabilities with detailed analysis;
  \item Study state-of-the-art generic mitigations on browsers;
  \item Perform a detailed study on a real-world full-chain exploit.
\squishend

\PP{Non-goals of this SoK.}
The main focus of this study is \emph{web browser} security,
concerning the security of its own vulnerabilities.
We do not consider other web-related security problems,
like \emph{web} or \emph{web server} security issues
such as Cross-Site Scripting (XSS), SQL Injection, etc.
Note that although 
\emph{Universal Cross-Site Scripting (UXSS)}~\cite{google-uxss-study} 
sounds similar to XSS,
it commonly originates
from problems in the browser's implementation and design,
so it is considered \emph{web browser} security (\autoref{s:uxss}).

%% file: architecture.tex
\section{Modern Browser Architecture}
\label{s:arch}

This section
provides a unified model of each web browser
that enables an objective comparison
of their approaches. %to security. 

\subsection{Overview}
\label{s:arch:overview}
Modern web browsers adopt the principle of least privilege
by using an OS process
as a protection domain.
By using the process domain,
each web browser can be
described using three types of processes,
namely, 
a \emph{browser} process (marked in green), 
\emph{renderer} processes (marked in magenta),
and task-specific processes (marked in yellow),
as shown in~\autoref{fig:arch}.

\PP{Browser process.}
When a web browser launches,
the browser process runs with the same privilege level as the user
(\ie, a higher privilege)
and passes a sandbox profile to the OS
to restrict the privileges of other processes to be spawned with
(\ie, a lower privilege). 
It manages all child processes (\eg, renderer)
and is the only process
that directly interacts with users
via system calls (and a user interface).

\PP{Renderer process.}
This process is responsible
for parsing and rendering
the untrusted web content.
The ever-growing kinds
of data served on
the web have caused the renderer process
to include
a wide variety of components,
such as media parsers,
DOM and JS engines.
Since they are major sources of browser bugs,
they are confined in
a restrictive sandbox (see~\autoref{s:sandbox}).
The renderer processes
are typically spawned per browser tab
or per web page origin.
The isolation policy of each renderer
varies by security policy or features (\eg, site isolation) of each web browser,
available resources at runtime (\eg, low memory in mobile),
or even user configuration.

\PP{Other process}
A modern browser's architecture is highly modular.
This modular design enables browsers
to have different privilege levels based on
the process's role.
Services that interact with external drivers
(\eg, networking or GPU processes)
are isolated as a separate process,
which enables
more restrictive sandboxing
for the processes that don't require such access
like renderer process.
Web browsers also commonly put extensions
and plugins in separate processes.
This protects plugins
that are at
a higher privilege level from malicious web content,
and  protects
browsers from being hijacked in
the case of
a malicious plugin.

\PP{Inter-Process Communication (IPC)}
Since these processes cannot directly access each other's memory,
they always communicate via IPC channels
provided by the OS,
and communications are usually mediated by the browser 
(\emph{broker}) process.
In other words,
the browser process works as a reference monitor
that restricts direct accesses
to important data or high-privileged operations
(\eg, cookies or system calls)
from other processes.
Thanks to this multi-process architecture,
an attack is always initiated from
a low privileged process like a renderer process,
and the attacker's goal is
to break into the browser process running as a user's privilege.
At the same time,
it makes it possible to recover from crashes caused by
a benign bug in the renderer process,
making the browser resilient against stability issues. \looseness=-1

\PP{Same-Origin Policy (SOP)}
In reality,
websites consist of contents from numerous sources with varying origins,
\eg, using CDN for common \js libraries,
embedding external sites via \emph{iframes},
or enabling a \emph{like} button from a social network.
The complex nature of websites
leads to numerous security policies
and unique features
of each web browser.
Based on the origin of each website~\cite{sop-MDN},
the browser process
and the renderer process
restrict
which resources (\eg, cookies)
a web page is allowed to interact with,
which is the same-origin policy (SOP)~\cite{sop-MDN}.

\subsection{Differences in Browsers}
\label{s:arch:diff}
\label{s:arch:jse}
The so-far discussed design is equally applied
to all four major browsers.
However,
as shown in~\autoref{fig:arch},
some implementation details differ depending on
the design of the browsers
and their underlying operating systems.
For example,
the GPU processes in Chrome and Safari are separated from
the renderer processes,
with a sandbox profile
that enables them to access
the platform 3D APIs~\cite{webkit-gpu-sb} (see~\autoref{s:sandbox}).
Also,
Chrome, Firefox, and Safari each has
a separate process to handle
the network service,
while the Chrome network service is placed 
outside the sandbox. 
Chrome team is currently implementing
the sandbox of its network service~\cite{chrome-networking-sb}.

\PP{Site isolation}
The sandbox mechanism can indeed protect browsers;
however,
with the discovery of universal cross-site scripting (UXSS) attacks,
it turned out that attackers could steal user data without
needing to escape the sandbox.
To address such attacks,
the Chrome team came up with \emph{Site Isolation}~\cite{reis2019site}
to further separate different site origins into
different processes.
It creates
a dedicated process for each site origin,
so that there is no implict sharing among different origins. 
Site isolation is
an effective measure
to address UXSS,
but it is also beneficial
for preventing hardware-based transient execution 
attacks~\cite{kocher2019spectre, lipp2018meltdown}.
Firefox has
a similar project named {\it Fission}~\cite{fission},
and it is shipped in Firefox 88 Beta~\cite{ff88beta-news}. 

\PP{JavaScript engines.}
\js engines
are the core of modern browsers,
which convert \js code into machine code. 
Major browsers use just-in-time (JIT) 
compilation~\cite{v8}~\cite{spm}~\cite{jsc}
to speed up
the code execution.
% %
Also,
JIT compilers model the result
and side-effects of all operations
and run various analysis passes
to optimize the code.
If any of these goes wrong,
native code with memory corruption issues
can be emitted
and executed,
which can lead to severe security 
implications~\cite{jit-exploitation,pzero-jit}.
While each engine has different implementations,
they share similar design principles
and have common attack surfaces
(\autoref{s:vuls:jit}).
Therefore, attackers can build
generic attack primitives which 
work across different engines,
such as \emph{fakeobj} and \emph{addrof} 
primitives~\cite{phrack-addrof,inline-caching}
and \emph{element kind transitions}~\cite{phrack-jit,element-kinds}.
\js engines are being used outside browsers as well 
(\eg Node.js uses V8),
amplifying the impact of security bugs in \js engines.
We discuss issues caused by homogeneous browser engines
in~\autoref{s:patch-gap}.
%}

\PP{Rendering engines.}
Rendering engines are responsible for interpreting resources
and rendering webpages.
Each of the four major browsers has its own rendering engine:
Safari uses WebKit;
Chrome uses Blink (forked from WebKit);
Firefox uses Gecko;
Edge uses Blink (replacing EdgeHTML).
\emph{Web Standards}~\cite{w3c,whatwg} 
serve as baseline specifications
and references for browser vendors 
to implement their rendering engines.
Since \emph{Web Standards} continuously evolve
with new features,
there are rapid changes in rendering engines, \ie,
implementing new features or dropping deprecated ones.
Due to different decision process 
and  implementation strategy, 
the feature sets implemented
in the rendering engines in different browsers
are quite different~\cite{caniuse-dot-com}, 
resulting in different attack surfaces~\cite{yason2015understanding}.
We discuss the attack surfaces in~\autoref{s:vuls}.

\begin{table*}[t]
  \centering
  \footnotesize
  \begin{tabular}{>{\centering\arraybackslash}m{420pt}}
    \textbf{Legend$\dagger$:}
    \textbf{GMP:} Gecko Media Plugin,
    \textbf{UNT:} Untrusted,
    \textbf{MED:} Medium,
    \textbf{LMT:} Limited,
    \textbf{LKD:} Lockdown,
    \textbf{NAD:} Non-Admin,
    \textbf{LTU:} Limited User,
    \textbf{IZG:} Inherited from Zygote,
    \textbf{BSC:} seccomp-BPF + Sandboxed IPC
  \end{tabular}
  \includegraphics[width=.95\textwidth]{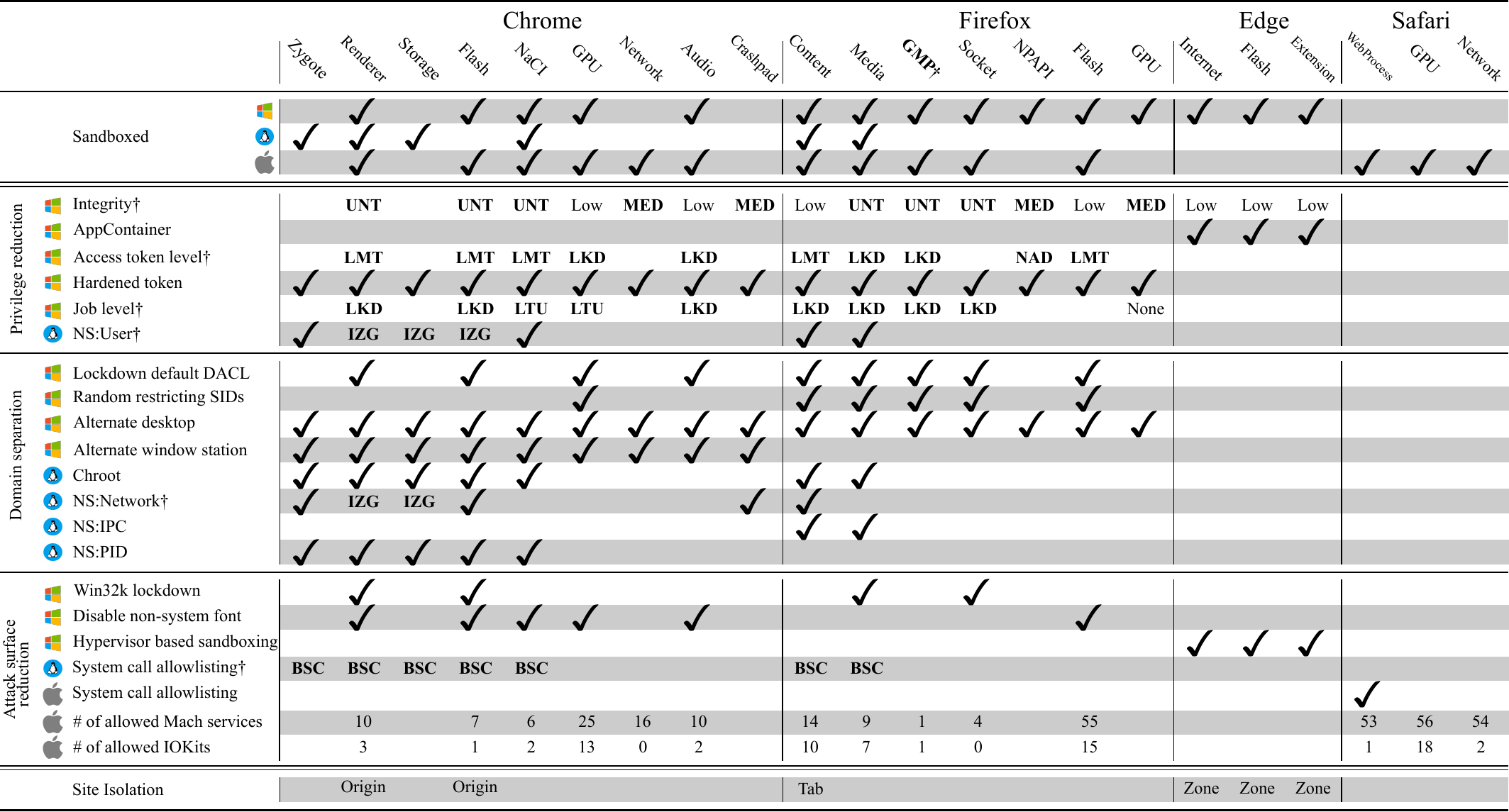}
  \caption{Sandbox comparison (Chrome, Firefox, Edge, Safari on Windows,
    Linux, MacOS)}
  \label{t:sandbox}
  \vspace{-13pt}
\end{table*}

\subsection{Variances in Sandbox Schemes}
\label{s:sandbox}

The sandbox restricts
the program execution from 
deviating from its intended mission.
However,
the underlying technology
and architecture for building 
a sandboxed environment significantly differs among OSes.
To examine the internals of sandbox implementations, we 
1) audit the source code of browsers, 
2) monitor the behavior of the sandbox APIs, and 
3) analyze the predefined sandbox policy file (\eg, \safari browser's configuration).
We summarize our findings in~\autoref{t:sandbox}.

\PP{Categorizing sandbox primitives.}
In~\autoref{t:sandbox}, we categorize sandboxing primitives 
into three categories based on their roles:
\textit{a) privilege reduction}
applies more restricted privileges to
the sandboxed processes using
the permission system of platforms such as DAC/MAC;
\textit{b) domain separation}
allocates a separated space of resources 
that a sandboxed process will have access to; 
\textit{c) attack surface reduction}
limits accesses to system services, kernel or device drivers.

\PP{Browser-specific characteristics.}
Browser vendors utilize different primitives depending on
the given constraints
(\eg, available memory).
For example,
Site Isolation prevents RCE exploits to be transformed
into UXSS or sandbox escapes by putting an origin-wise,
process-level security boundary
between a compromised renderer
and privileged web pages~\cite{gong2015pwn, gong2016pwn}.

\PP{OS-specific behaviors.}
We also compare the sandbox features
from different OSes, namely, Windows, Linux and macOS.

% P1. Windows
\PPi{Windows}
Windows restricts each process
by using a security token~\cite{chrome-sandbox}.
Similar to the capability-based model,
a process obtaining a certain token level
can access privileged resources
with proper security descriptor level.
For example,
the renderer process runs with a \emph{low integrity} token level,
and the broker process
runs with a \emph{medium integrity} token level,
so any write accesses from a renderer process to the broker process
will be restricted by default.\looseness=-1

However,
there is no unified protocol
for fine-grained access control.
To resolve this,
\chrome and \ff support fine-grained rulesets
using their own IPC mechanisms
and binary-level code patches on
resource-related functions~\cite{chrome-sandbox}.
%
% In addition,
Microsoft introduced {AppContainer} in Windows 8 to 
enforce more fine-grained access control
to resources by adding
a notion of {capabilities} attached to process tokens.
Edge created
a sandbox based on AppContainer~\cite{edge-sandbox-tuning}.
Starting from
the \emph{deny-by-default} policy,
Edge created
a set of capabilities for required system resources.
Chrome is also experimenting 
with an AppContainer-based sandbox~\cite{chrome-networking-sb}.
%
% Domain separation
Browsers also utilize various features
for mitigating sandbox escape. 
For example,
\emph{alternate desktop} and 
\emph{alternate window station} can be used to 
mitigate UI-based attacks such as Shatter~\cite{shatter}; 
\emph{lockdown default DACL}~\cite{lockdown-dacl} and \emph{Random Restricting SIDs}~\cite{random-restricting-sid} 
were introduced
to enforce more restricted DACLs,
so that compromised sandboxed processes 
cannot access other sandboxed processes.

% P2. Linux
\PPi{Linux}   
Unlike on Windows,
the Linux sandbox is mainly based on \emph{seccomp},
\emph{chroot} and \emph{namespace}. 
%\footnote{https://man7.org/linux/man-pages/man2/seccomp.2.html} 
%\footnote{https://man7.org/linux/man-pages/man2/chroot.2.html}
%
First,
\emph{seccomp} is
a standard system call filter
based on the eBPF language.
Since the default \emph{seccomp} configuration is overly tight,
browsers define their own filtering rules.
For example,
\chrome applies its custom \emph{seccomp} rules to all processes
except the broker process,
and the detailed rules vary for each process.
Second,
to restrict file access,
Linux-based browser sandboxes utilize \emph{chroot} jailing.
Once a process is confined with \emph{chroot},
no upper hierarchy of the file system is reachable.
For example,
\ff applies \emph{chroot} jailing to all renderers
and only allows them to access specific files
based on file descriptors obtained from the broker process. 
Also,
browsers use \emph{namespaces}~\cite{namespaces-linux}
to create separated spaces
for various resources,
such as user, networking, and IPC.
For example,
creating and joining a user {namespace} enables
a sandboxed process to be in a separate UID and GID,
effectively disabling access to other unsandboxed processes. \looseness=-1

% P3. OSX
\PPi{macOS}
While Windows and Linux support various types of sandboxing primitives,
\macos supports a specifically formatted 
\textit{sandbox profile} (\cc{.sb}) \cite{seatbelt-reversing} 
to describe the sandbox policy for a given process. 
Typically,
the file provides
an allowlist of absolute file paths that are
allowed to be accessed
and blocks all other file accesses by default.
The profile also defines
the capability of accessing other resources 
such as network and shared memory, 
and supports systemcall-based filtering like Linux's \emph{seccomp},
though it is only deployed on \safari.

\PP{Mobile platforms.}
Since a process-based sandbox uses
a non-trivial amount of memory,
mobile platforms
introduce subtle differences in sandbox policies
or disable them depending on the available resources.
For example, on Android, Site Isolation in \chrome
is enabled only when the device has enough memory (>1.9GB),
and
the user need to enter passwords on the website~\cite{site-isolation-android}.
On iOS,
\safari uses sandbox rules that are different from macOS 
because different system services
and IOKit drivers are exposed on mobile.
Due to such differences,
some exploits may work only on mobile platforms~\cite{android-bitmap-exploit}.

\begin{figure*}[t]
  \centering \includegraphics[width=0.95\textwidth]{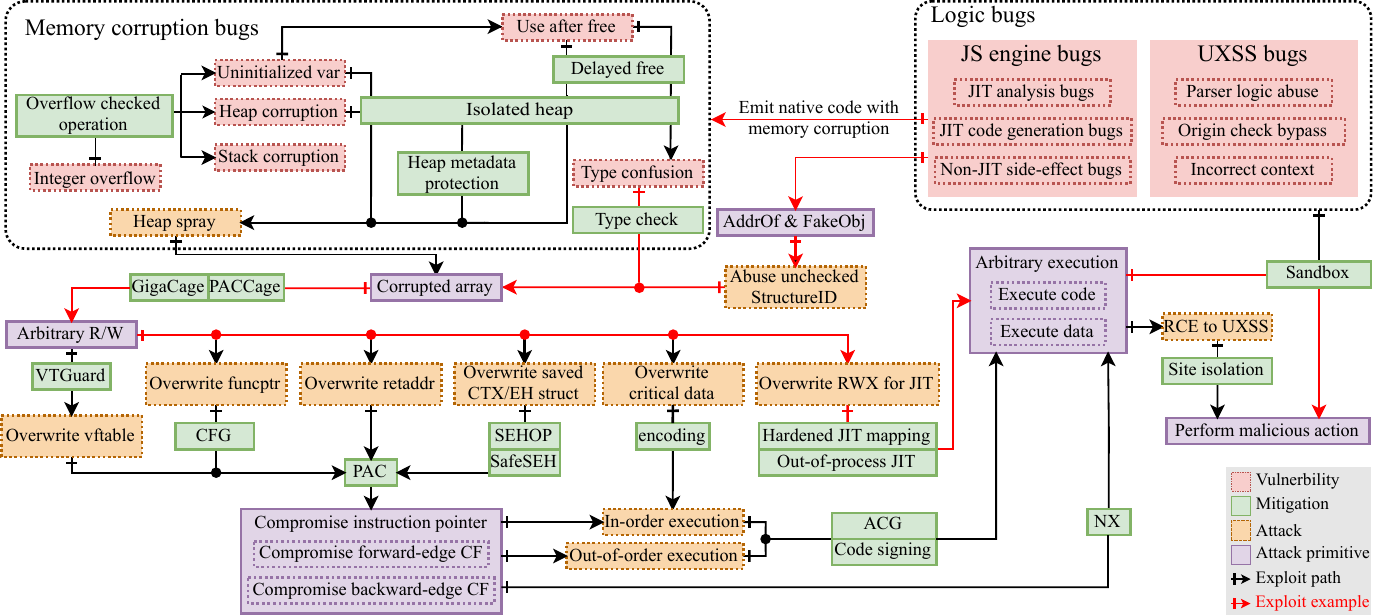}
  \caption{
    Browser exploitation scenarios and bug classification.
    We mainly focus on browser security-specific issues and omit basic
    software attack/defense techniques such as ROP, NX and ASLR.
    The \textbf{Exploit example} arrows depict the exploit path 
    described in~\autoref{s:pwn2own}.
    } 
  \label{fig:main}
  \label{fig:exploit}
  \vspace{-13pt}
\end{figure*}

\subsection{Exploiting Browsers}
\label{s:arch:exploit}

The goal of browser exploitation
is to steal sensitive data from its user
or to install malware for further action.
Attackers can execute attacks like UXSS
to directly steal data, or
get a code execution first
and then try to escape the sandbox.
An attacker might attempt to
gain a system's privilege
and escape the sandbox
by attacking the kernel,
which is out of the scope of this paper.
Thanks to various mitigation schemes 
(see~\autoref{s:vuls},\autoref{s:mitigation}),
an attacker should chain
multiple bugs
(\eg, 4 bugs until sandbox escape~\cite{jin:pwn2own2020-safari})
together
to gain an arbitrary execution.
Even after the control-flow is hijacked,
since the renderer process runs
inside the sandbox, 
the attacker should find another set of bugs
in the broker process
to escape the sandbox.
Depending on exploits available in one's arsenal,
an attacker often
attempts to exploit a bug in
the system services, drivers, or
the kernel
instead of the broker process
to break out of the sandbox~\cite{shell-on-earth}.

\begin{comment}
A full-chain browser exploit usually starts with
a code execution exploit against a renderer because
the attack surface is concentrated in it.
When exploiting the side effect modeling bug of the JIT compiler,
we can make a JIT-compiled function with incorrect assumptions in emitted code
leading to exploitable memory corruption such as type confusion
(\eg accessing an array with double as an array with double/object pointer and vice versa)
and result in addrof primitive and fakeobj primitive at the end.
We can use this primitive to fake an array object
to get arbitrary read-write
and overwrite things to finally execute a code.
Mitigations like Structure ID Randomization make creating fake objects harder,
and Address Space Isolation type of mitigations prevents crafted or corrupted array objects from accessing full address space.
As the last line of defense,
CFI mitigations and overwrite protection type of mitigation aim to prevent arbitrary code execution.
However,
they're still bypassable, for example, by overwriting RWX JIT pages for WASM. 
Finally,
an attacker needs to break out of the sandbox by exploiting the browser or kernel.
Connection of various vulnerability classes, exploit techniques, primitives, mitigations are illustrated in \autoref{fig:main}.
\end{comment}

%% file: sok.tex
\section{Browser Vulnerabilities and Mitigations}
\label{s:overview:class}
\label{s:vuls}
In this section, we first perform a measurement study 
on publicly reported browser bugs in the past decade 
to reason about the trends and then present dominant types 
of vulnerabilities (\eg, \js engine bugs) and 
their corresponding mitigations deployed by the vendors.

\input{trend.tex}

\PP{Bug types included in this section.}
Based on the trend of bugs presented in~\autoref{fig:exp-bugs},
in this section, we mainly discuss the trending types of bugs,
namely:
1) Parser bugs (\autoref{s:data-parsing-bug}),
2) DOM bugs (\autoref{s:dom-bugs}),
3) \js Engine bugs (\autoref{s:vuls:jit}),
4) SOP bypass and UXSS bugs (\autoref{s:uxss}).

\subsection{Parser Bugs}
\label{s:data-parsing-bug}
%\JL{better section name?}
Parsers often suffer from memory corruption bugs; 
there is no exception for parsers in browsers.
In web browsers, the majority of parser bugs have been found 
in media parsers or network protocol parsers. 
As shown in~\autoref{fig:ff-ch-comp},
\emph{Renderer (Media)} takes a large share.
These bugs are easier to exploit in the 
renderer process since they can be utilized to corrupt 
JS objects and create more powerful exploit primitives.

\PP{Current status.}
After the hardening of heap allocators 
(\autoref{s:hardened-allocators}), 
such exploits were made much more difficult or infeasible, 
mainly because of the compartmentalization of JS objects on the heap. 
Also, large-scale fuzzers like ClusterFuzz~\cite{clusterfuzz} 
have discovered many parser bugs. 
Browser vendors are working on sandboxing networking code 
and rewriting browser code using memory-safe languages like Rust~\cite{oxid}. 
As a result, these bugs have become scarce and harder to exploit.
Still, there are multiple dependencies of third-party libraries
when it come to parsing data,
so tight control of security updates are needed.

\subsection{DOM Bugs}
\label{s:dom-bugs}
DOM bugs were popular targets for attackers; 
according to~\autoref{fig:exp-bugs}, 
the majority of exploited bugs in 2014 were DOM bugs. 
Since most of them were
UAF bugs,
mitigations have been deployed to reduce the exploitability of
them, 
such as \emph{isolated heap} and 
\emph{delayed free}~\autoref{s:hardened-allocators}.

\PP{Current status.}
While fuzzers continue to identify
new DOM bugs~\cite{clusterfuzz, xu2020freedom, domato},
as shown in~\autoref{fig:exp-bugs},
recent known in-the-wild full-chain exploits tend to use bugs 
in other components due to the increased difficulty 
of exploiting DOM bugs.

\lesson{UAF mitigations are effective towards reducing DOM bug exploits}
{
Since DOM bugs mostly rely on UAF problems, 
they have been mostly mitigated by UAF mitigations.
Generic exploitation techniques
relying on pointer type confusion
have become infeasible
since heaps are isolated by object types,
and there are no publicly known alternative techniques.
As a result, exploiting DOM bugs 
is no longer a preferred way to compromise renderers.
}

\subsection{JS Engine Bugs}
\label{s:vuls:jit}
\label{s:mitigation:primElim}
\label{s:overwrite-protection}
In recent browser exploits, 
JS engine bugs are one of the most 
popular targets of browser exploits, 
especially optimization bugs.
At least 34\% of exploited bugs (\autoref{fig:exp-bugs})
have utilized JS engine bugs 
for compromising the renderer process, 
which is usually the first step for full-chain browser exploits.
JS engine bugs can be utilized to easily generate powerful exploit primitives 
like \cc{addrof} (to leak the address of any JS object) 
and \cc{fakeobj} (to access an arbitrary address as an object).
As mentioned in~\autoref{s:arch:jse}, 
JIT compilers in JS engines use speculative optimization.
Bugs in these optimizations are far more critical 
than conventional memory safety bugs such as 
use-after-free or buffer-overflow, 
as they are hard to mitigate but provide
powerful exploitation primitives to attackers.
On a high level, JS engine bugs can be mainly divided into four categories: 
\squishlist
    \item \textit{JIT analysis bugs:} 
    Bugs in the analysis process or models of the JIT compiler. 
    Such bugs have the highest exploitability and impact. 
    \item \textit{JIT code mutation/generation bugs:}
    Bugs in the process of manipulating JIT graphs 
    or emitting code. They often result in an outright 
    unexploitable crash.
    \item \textit{Non-JIT side-effect bugs:}
    Side-effect bugs in \js built-in functions, 
    which are mostly related to fast-paths. 
    \item \textit{Non-JIT traditional memory corruption bugs:}
    Other memory corruption bugs that don't fall into the categories above.
\squishend

We examined the 45 exploited bugs in~\autoref{fig:exp-bugs};
there are 13 JIT analysis bugs, 9 non-JIT side-effect bugs and 
11 traditional memory corruption bugs, but there are no 
JIT code mutation/generation bugs. 
We suspect that this is because generation bugs are 
hard to exploit.
Most of the bugs in the JIT compiler are logic bugs.
Since it is a compiler infrastructure, 
logic errors can be amplified to memory safety errors 
in JIT-compiled code. Therefore, 
it is hard to make a general mitigation for JIT bugs. 
Here, we introduce three major categories of defenses:
primitive elimination, overwrite protection and jit-based code-reuse mitigations. \looseness=-1

\PP{Primitive elimination}
Primitive elimination techniques aim to prevent attackers from 
1) converting vulnerabilities to exploit primitives and 
2) escalating exploit primitives to stronger 
ones\footnote{For example, to construct reliable and 
stable read/write primitives,
an attacker can leverage the \emph{addrof} and \emph{fakeobj} 
primitives to fake an \texttt{ArrayBuffer} object 
with a fully controlled backing store pointer, 
which is an escalation of primitives.}.

\PPi{a) Object shape authentication.}
This type of mitigation aims to prevent attackers 
from crafting a valid object using the \emph{fakeobj} primitive.
    For example, in JavaScriptCore, StructureID Randomization 
    encodes the StructureID with seven random entropy bits, 
    which makes it hard for the attacker to 
    guess~\cite{webkit-structureid, pzero-jit}.  
    Since StructureID indicates the 
    \cc{type} and \cc{shape} of the JS object,
    incorrectly guessing the StructureID will lead to 
    invalid \texttt{shape}, 
    and accessing it
    will ultimately crash the process~\cite{structure-id-bypass}.

\PPi{b) Address space isolation.}
    This category of mitigations provides isolation of 
    different objects to prevent objects from being faked or overwritten. 
    GigaCage~\cite{fsecure-gigacage} is a 4GB virtual memory
    region that separates different objects into 
    different heaps (or \texttt{HeapKinds}). 
    The key idea is to prevent memory access across different heaps 
    and use the relative offset from the heap base address
    to locate a GigaCaged object, 
    instead of using absolute addresses.
    As such, even if a pointer is corrupted 
    it cannot point to anything outside its original heap.
    PACCage~\cite{pzero-jit} is applied to protect 
    the backing store buffer pointers of \texttt{TypedArray} 
    with Pointer Authentication Codes (PAC) on top of GigaCage
    to enhance the security even further.
    Chrome V8 Heap Sandbox~\cite{v8-heap-sandbox}, 
    which is  experimental, 
    has goals similar to GigaCage,
    but it tries to protect external pointers 
    using a separate pointer table, 
    so that attackers cannot create arbitrary values for external pointers.

\PP{Overwrite protection}
Overwrite protections are standard protection mechanisms to prevent attackers 
from introducing arbitrary executable code, 
which can be seen as the last line of defense 
in the context of browser exploits.
They mainly include four mechanisms: 
W $\oplus$ X~\cite{gawlik2018sok}, 
hardened JIT mapping~\cite{ios-jit-harden}, 
fast permission switch~\cite{ios-jit-harden,intel-mpk}, 
and out-of-process JIT~\cite{edge-acg}.

\PPi{a) W $\oplus$ X} 
W $\oplus$ X~\cite{gawlik2018sok} is an important security principle 
that enforces memory to be 
either executable but not writable
or writable but not executable. 
This mitigation made traditional shellcode injection attacks 
completely obsolete and provided the foundation for 
many other protection techniques~\cite{abadi2009control, zhang2013practical}. 
Surprisingly, JIT code pages are often exempt from 
this basic mitigation and mapped as \cc{rwx} for performance reasons~\cite{gawlik2018sok}.

\PPi{b) Execute only memory}
iOS 10 on ARMv8 devices landed
hardware support for \emph{execute-only memory} (XOM)~\cite{ios-jit-harden},
enabling JIT-compiled code to
contain secret data as an immediate value.
Safari
utilizes XOM
to hide the address of the writable-executable mapping
from attackers,
by introducing an execute-only \cc{jit\_memcpy} function
that has the base address of JIT mapping inside.
This makes arbitrary read/write insufficient 
for the JIT code page overwrite,
and forces attackers to take an alternative path 
\eg, hijacking control-flow to call \cc{jit\_memcpy}.

\PPi{c) Fast permission switch: APRR \& MPK}
Hardware support for fast permission switching was introduced 
to reduce the overhead of switching page permissions using \cc{mprotect()}.
Since iOS 11 on ARMv8 devices,
APRR~\cite{ios-jit-harden} was deployed to enable per-thread permissions 
by mapping page permissions (\cc{r,w,x}) to eight dedicated registers 
that indicate the actual page permission of the thread.
Similarly, Intel MPK~\cite{intel-mpk} adds a separate 4-bit integer per page,
% which is mapped to one \cc{PKRU register} that 
to enforce two additional protections: 
\emph{disable access} and \emph{disable write}. 
Consequently, the JIT region will always be 
\cc{r-x}, and only the write operations from a dedicated data-copying thread
are allowed by invoking an \cc{unlock} function,
which changes the permission to \texttt{rw-} only for the target thread.

\PPi{d) Out-of-Process JIT} 
On Windows, mitigations like Arbitrary Code Guard (ACG) 
ensure that
a process can only map \textit{signed} code into its memory. 
However, browsers heavily use JIT compilers for performance purposes,
which generate unsigned native code in a content process.  
Out-of-process JIT~\cite{edge-acg} was introduced
to enable ACG with JIT compilers.
Consequently, the JIT functionality
was moved to a separate process that runs in its own sandbox,
and it is responsible for compiling JS code and mapping it into 
the process. 
As such, the content process is never allowed to 
map or modify its own JIT code pages.

\PP{JIT-based code-reuse mitigations.}
\emph{JIT spray}~\cite{warpjit-guide} is a technique that
injects a vast amount of attacker-controlled 
JIT code (marked as \cc{executable})
into a predictable address in the memory to bypass ASLR/DEP, 
similar to \emph{Heap spray}~\cite{heap-spray}.
To mitigate JIT spray, browsers
put a size limit on JIT code
and
switched to 64-bit platforms 
with high-entropy ASLR, 
which made JIT-spray infeasible.
Still, it is possible to utilize the JIT code gadgets 
if their addresses are known to the attacker.
Such attacks are called \emph{JIT-based code reuse attacks} (JCRAs).
Here, we briefly summarize mitigations for such attacks.

\PPi{a) Controlled bytes elimination}
JCRAs have a fundamental assumption that 
with control of immediate operands and specific opcodes,
%used by the JIT compiler to generate the machine code, 
an attacker can control the generated JITed code in heap memory.
Therefore, mitigations were proposed to eliminate 
the predictability of attacker-controlled bytes, 
such as obfuscating large constants~\cite{athanasakis2015devil, lian2017call},
permuting the register allocation of 
immediate operands and local variables~\cite{wei2010secure, lian2017call}
and jumbling the instructions 
in a function's call frame~\cite{lian2017call, wei2010secure}.

\PPi{b) Internal randomization} 
Attackers can also leverage the relative location 
of instructions with each other or 
predictable offsets from the base address. 
Some of the mitigations aim to diversify the JIT code layout, including:
randomizing the relative offsets between different pairs of 
instructions~\cite{lian2017call, homescu2013librando, wu2012rim},
and inserting free space randomly before the first unit 
of code~\cite{lian2017call,gawlik2018sok}.

\PP{Current status.}
% JS engine
While there are some trials to prevent certain types of errors 
(\autoref{s:mitigation:sc}), it's hard to cover all of them.
As a result, mitigations in JS engines 
focus on eliminating attack primitives. 
Recently, the Edge team added a new security feature called 
Super Duper Secure Mode (SDSM)~\cite{sdsm1,sdsm2},
which basically disables JIT compilation. 
Users can choose to disable JIT on websites 
that are less frequently visited.
While sacrificing some performance,
it is a good approach for reducing attack surfaces.
% JIT-based code-reuse
For JCRAs,
although multiple mitigations have been introduced,
they are still 
viable~\cite{chrome-constant-blinding-bypass,gawlik2018sok}
since vendors did not put many resources into implementing or 
maintaining mitigations.

\lesson{Mitigating JS engine bugs is difficult}{
JavaScript engine bugs, especially JIT compiler bugs, 
are very powerful since the attacker 
can emit code with memory corruption issues. 
Many mitigations aim to prevent 
escalation of exploit primitives 
because it is hard to mitigate logic bugs in general.
Therefore,
vendors often deploy mitigations that aim to break exploit paths,
and enhance them continuously to prevent future attacks.
}

\subsection{SOP-Bypass and UXSS Bugs}
\label{s:uxss}
\emph{Same origin policy} (SOP)~\cite{sop-MDN} is enforced by web browsers to 
keep a security boundary between different origins.
SOP-bypass bugs can be used to compromise SOP
to varying degrees, 
from leaking one bit to stealing full-page data.
UXSS bugs are 
the most powerful type of SOP-bypass bug that 
can be used to facilitate cross-origin \js code execution.
In UXSS attacks, the attacker can inject scripts to 
any affected context by exploiting bugs 
in web browsers~\cite{CVE-2021-1879,CVE-2021-34506} or 
third-party extensions~\cite{lastpass-uxss, ublock-uxss},
achieving the same effect as exploiting 
the XSS vulnerability in the target website.

\PP{Current status.}
\emph{Site Isolation}~\cite{reis2019site,siteisol} is one of the most 
significant mitigations against UXSS attacks.
Site Isolation enforces SOP at the process level, 
which made most existing UXSS bugs unexploitable.
The number of reported UXSS bugs was 
significantly reduced after site isolation 
was gradually applied after 2017, as shown 
in~\autoref{fig:bug-bounty}.
However,
UXSS vulnerabilities in third-party extensions
still exist;
multiple UXSS bugs have been found in popular 
extensions~\cite{lastpass-uxss, ublock-uxss}, 
which have enabled attackers to bypass site isolation and 
steal the user's credentials. 

\lesson{UXSS bugs are mostly mitigated by Site Isolation}{
Site isolation is an effective mitigation against UXSS bugs.
However,
only Chrome and Firefox have site isolation deployed,
since it requires a considerable amount of engineering effort 
(Appendix~\ref{s:difficulty-mitigations}).
}

\subsection{Summary}
Due to threat research and improved patch deployments, 
the impacts of 1-day exploits are reduced, 
and in-the-wild 0-day exploits get patched quickly
once they are caught.
However, offensive research is still much ahead of the vendors.
Although vendors are trying, 
they are consistently behind in this arms race.
Mitigations from vendors are mostly reactive, 
which means they are developed long after each wave of attacks.
By the time an attack surface is finally closed, 
attackers have already come up with a better exploit. 
It's a difficult task, but vendors should be 
more proactive and implement new features 
with security implications in mind,
\eg, studying potential 
new attacks before deploying new features.
%

%% file: trend.tex
\subsection{Trends of Browser Bugs}
\label{s:trends}
\PP{Data collection}
We study public CVEs
and vulnerability reports
for four major browsers:
1) routinely updated security advisories from 
'browser vendors~\cite{ff-sec-adv},
2) public issue trackers released by vendors%
~\cite{chr-bug-trac}~\cite{ff-bug-trac},
3) open source code repositories that have a convention of linking bug-fixing
commits to published vulnerabilities%
~\cite{ch}~\cite{chr-src}~\cite{v8-src}~\cite{ff-src},
4) CVE reports in the National Vulnerability Database (NVD)~\cite{nvd},
5) security bugs used in real-world exploits
such as bugs used in Pwn2Own~\cite{zdi},
and Google Project Zero reports~\cite{p0-bug-trac,p0-itw-bugs}.
\autoref{t:bug-numbers} summarizes the yield of our data collection efforts.

\begin{table}[t]
    \centering
    \footnotesize
    \setlength{\tabcolsep}{3pt}
        \begin{tabular}{l c c c c}
                \toprule
                & Firefox$\dagger$ & Chromium$\dagger$  &
                Safari/WebKit$\ddagger$ & Edge/IE$\ddagger$ \\
                \midrule

                (1) Total CVEs & 2190 & 2582 & 1436 & 2278\\

                (2) Collected CVEs & 2066 & 1912 & 1436 & 2278\\

                \midrule
                (3) Pwn2Own~\cite{zdi} & 14 & 9 & 37 & 31 \\ 

                (4) Google P0~\cite{p0-itw-bugs} & 7 & 22 & 8 & 22 \\

                \bottomrule

        \end{tabular}
    \caption{(1) The total number of CVEs in the NVD
    database~\cite{nvd}. (2) The number of collected bugs. 
    For open source browsers$\dagger$, we collected extended bug
    information from vendors' bug trackers
    ~\cite{ff-bug-trac}~\cite{chr-bug-trac}. 
    For those browsers, we ignored confidential bugs and bugs 
    not linked to bug tracker issues. 
    For closed source browsers$\ddagger$, we used NVD~\cite{nvd}
    as the sole source of CVE data. 
    (3) \& (4) are the sources used to collect data of {exploited bugs}.}
    \label{t:bug-numbers}
    \vspace{-13pt}
\end{table}

\begin{figure}[t!]
    \begin{center}
    \includegraphics[width=\columnwidth]{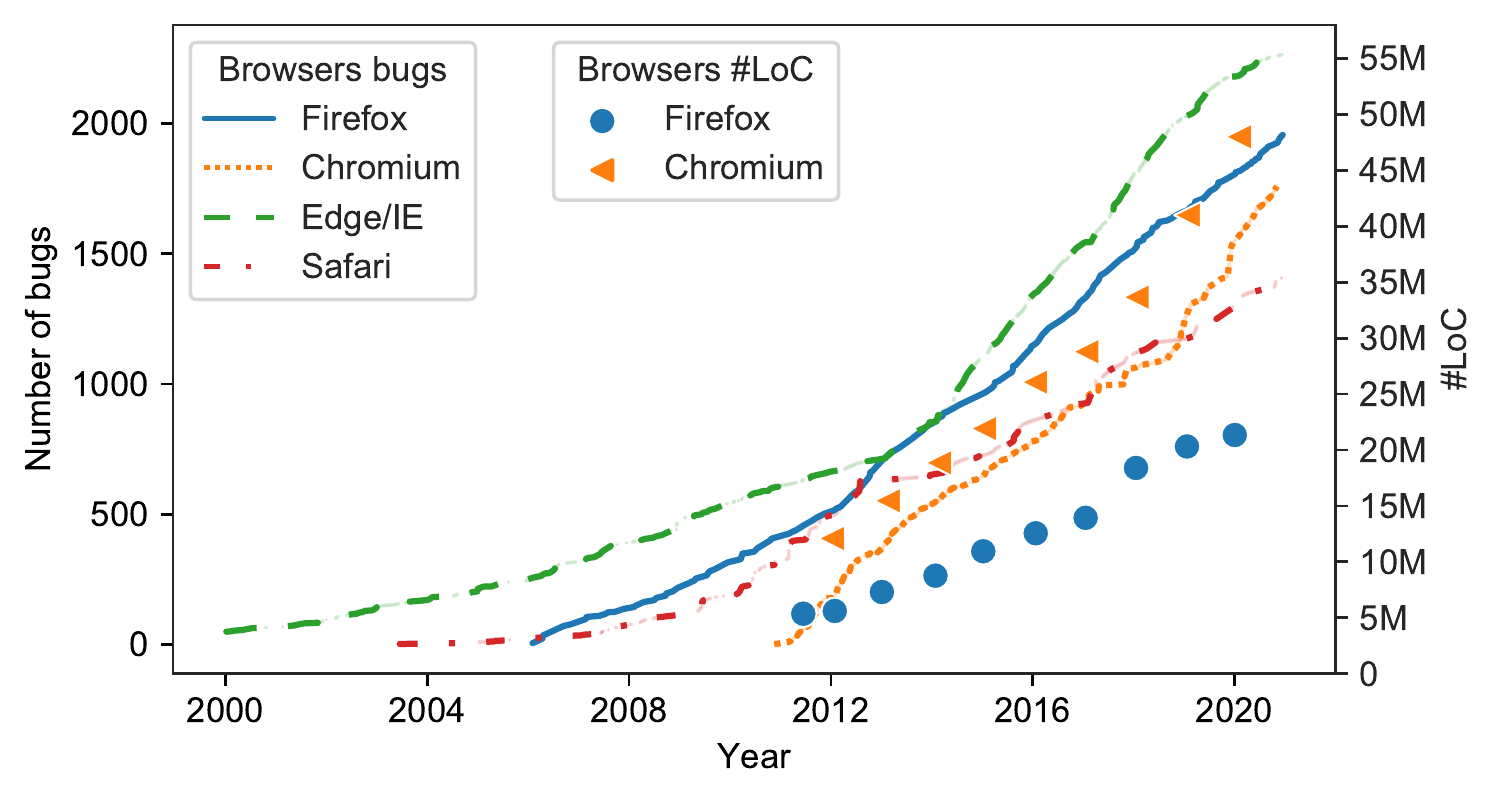}
    \end{center}
    \vspace{-10pt}
    \caption{
      Left y-axis: number of security bugs; Right y-axis: 
        LoC of two open-source browsers: Firefox and Chromium.
        LoC is based on the first major version bump each year. 
    }
    \label{fig:cve-numbers-browser-loc}
    \vspace{-13pt}
\end{figure}

\PP{Bugs and codebase size}
\autoref{fig:cve-numbers-browser-loc} shows the sharp increase 
of security bugs in all browsers, specifically starting after 2010.
We correlate this increase in bugs to the ever-growing codebase of browsers,
as new features are added constantly.
In addition, the advances in bug-finding techniques after 2010 played a
considerable role, which we highlight
in~\autoref{s:bug-finding-tool}.

\begin{figure*}[t]
    \centering
    \begin{subfigure}[b]{0.475\textwidth}
    \includegraphics[width=\textwidth]{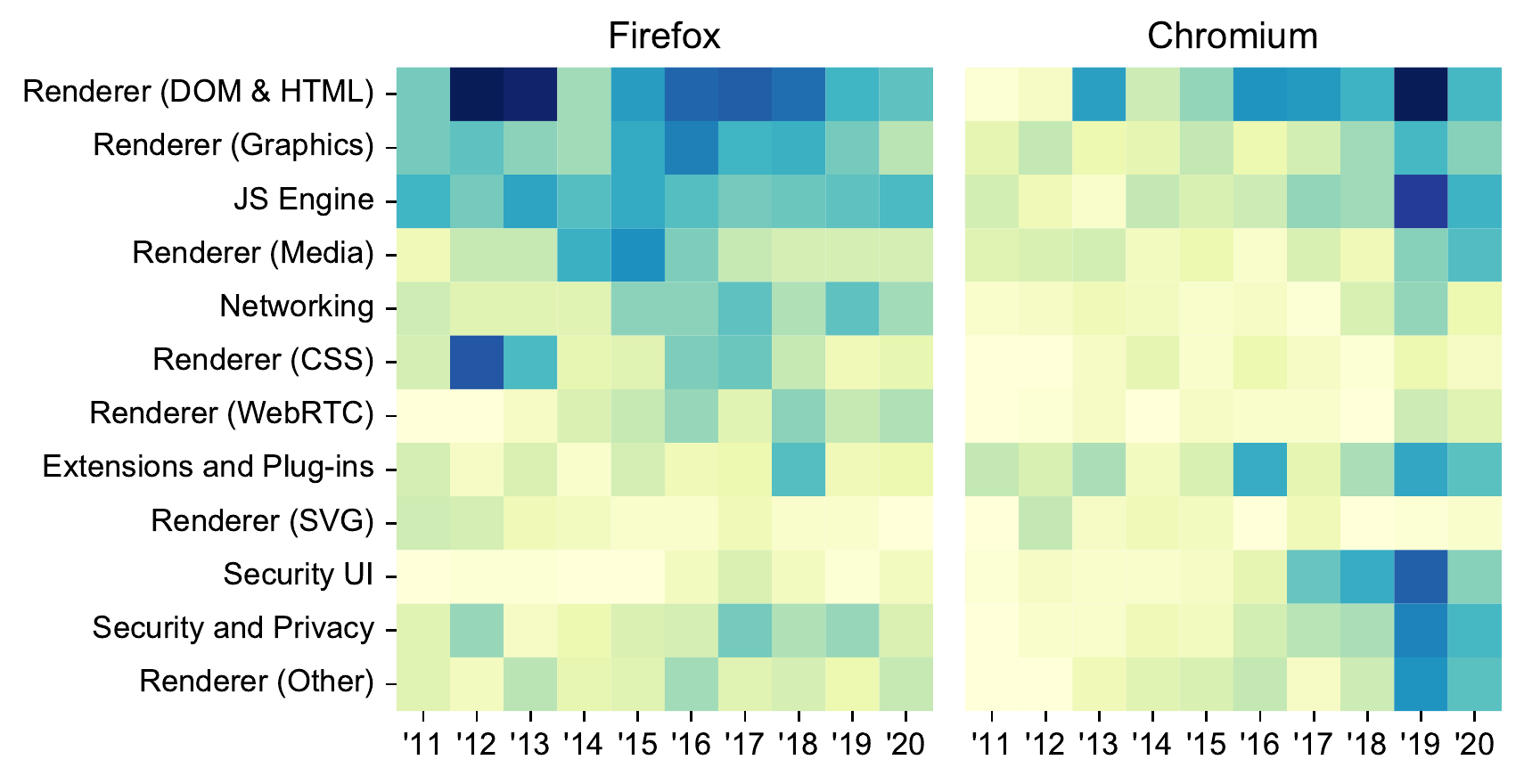}
    \caption{Browser Components}
    \label{fig:ff-ch-comp}
    \end{subfigure}
    \hfill
    \begin{subfigure}[b]{0.0508\textwidth}
    \includegraphics[width=\textwidth]{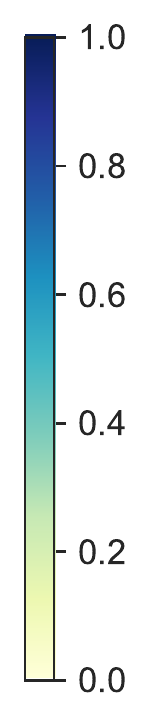}
    \caption*{}
    % \label{fig:ff-ch-class}
    \end{subfigure}
    \hfill
    \begin{subfigure}[b]{0.455\textwidth}
    \includegraphics[width=\textwidth]{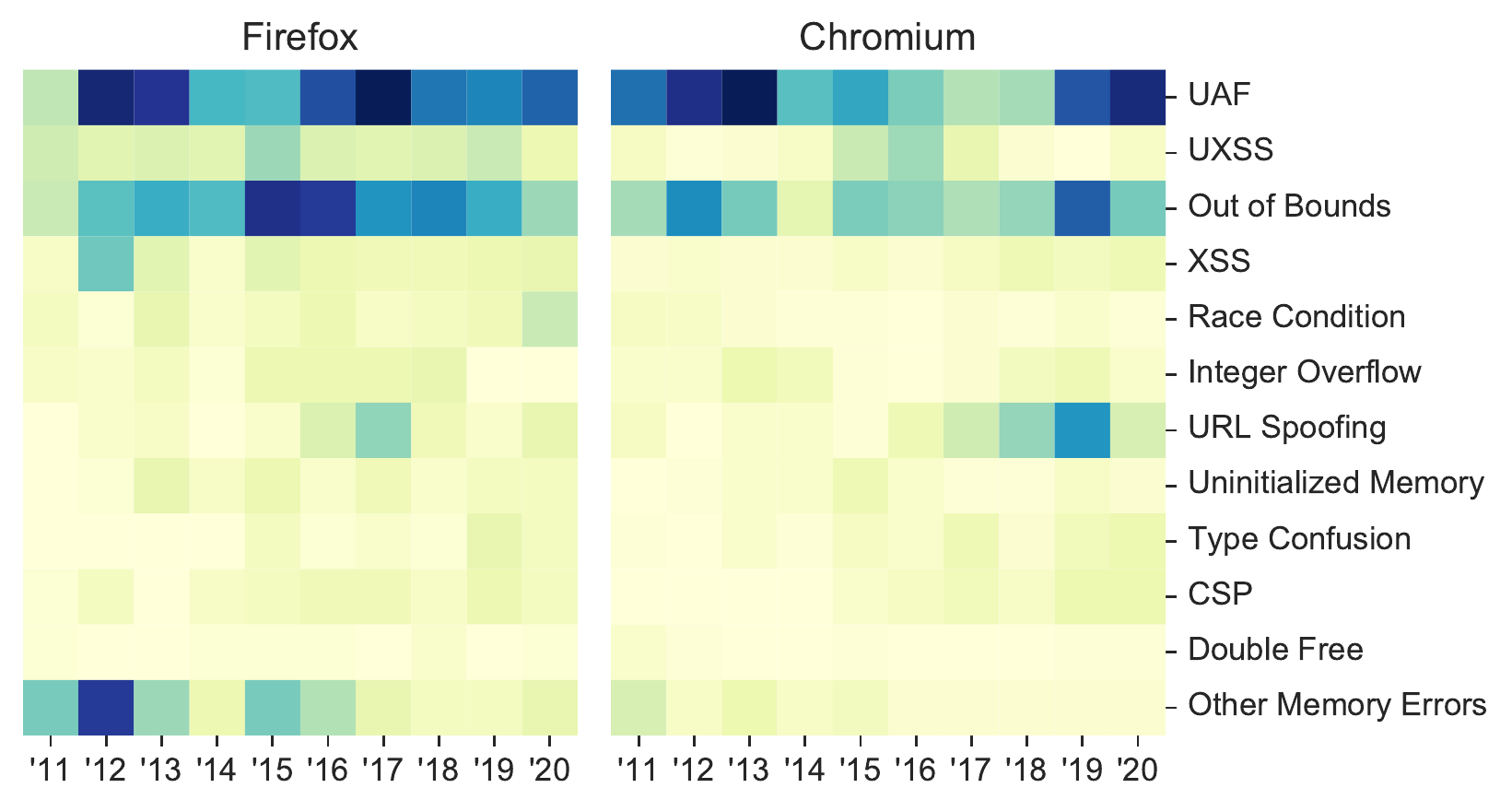}
    \caption{Bug Classes}
    \label{fig:ff-ch-class}
    \end{subfigure}
    \caption{
        Mapping of bugs to host browser components and bug classes 
        in Firefox and Chromium. The figure highlights the changing 
        nature of browsers' attack surfaces year-on-year. 
        The numbers in each figure are Min-Max scaled.
    }
    \label{fig:ff-ch-comp-class}
    \vspace{-12pt}
\end{figure*}

\PP{Dynamic attack vectors}
The enormous size and the continuously changing nature of browsers 
make the attack vectors change constantly.
%in a constant mutation.
%
For the open source browsers, Firefox and Chromium, we map bugs to their
respective host components and bug classes in~\autoref{fig:ff-ch-comp-class}. 
For both browsers, we use the developer's assigned flairs to map bugs 
to their host browser components, and we use keyword matching in 
bug descriptions to categorize their classes.

Renderer bugs are dominant in both Firefox and Chromium since renderers 
are the core of browsers.
The rise of URL spoofing bugs for Chromium since 2016
highlights the ease of finding bugs in previously unexplored areas. 
Memory bugs in general, and UAF bugs specifically, remain the
greatest common denominator bug class for both browsers.
Another general observation is the varying number of bugs across 
the two browsers along the years in both dimensions.
For example, for bug components, Chromium has more {DOM \& HTML} bugs 
recently, but the number of {DOM \& HTML} bugs is decreasing for \ff.
For bug class,
in 2019 most of
the bugs in Chromium were classified as UAF, OOB, 
and URL spoofing bugs,
but Firefox depicts a relatively uniform bug distribution across the years.
Thus, this discrepancy visualizes not only the changing attack vectors, 
but also the changing policies of triaging security bugs for different browsers.

Browsers' efforts against bugs can also be identified in the figures.
Chromium's Site Isolation~\cite{reis2019site,siteisol} as mitigation to 
UXSS bugs led to the apparent decrease of said bugs after Site Isolation 
was implemented in 2017 (\autoref{fig:ff-ch-class}).
Some parts remain as the main source of bugs such as the DOM \& HTML 
component, which we detail in \autoref{s:dom-bugs}.

\begin{figure*}
    \centering
    \begin{minipage}[t]{0.32\textwidth}
        \vspace{0pt}
        \centering
        \includegraphics[width=\columnwidth]{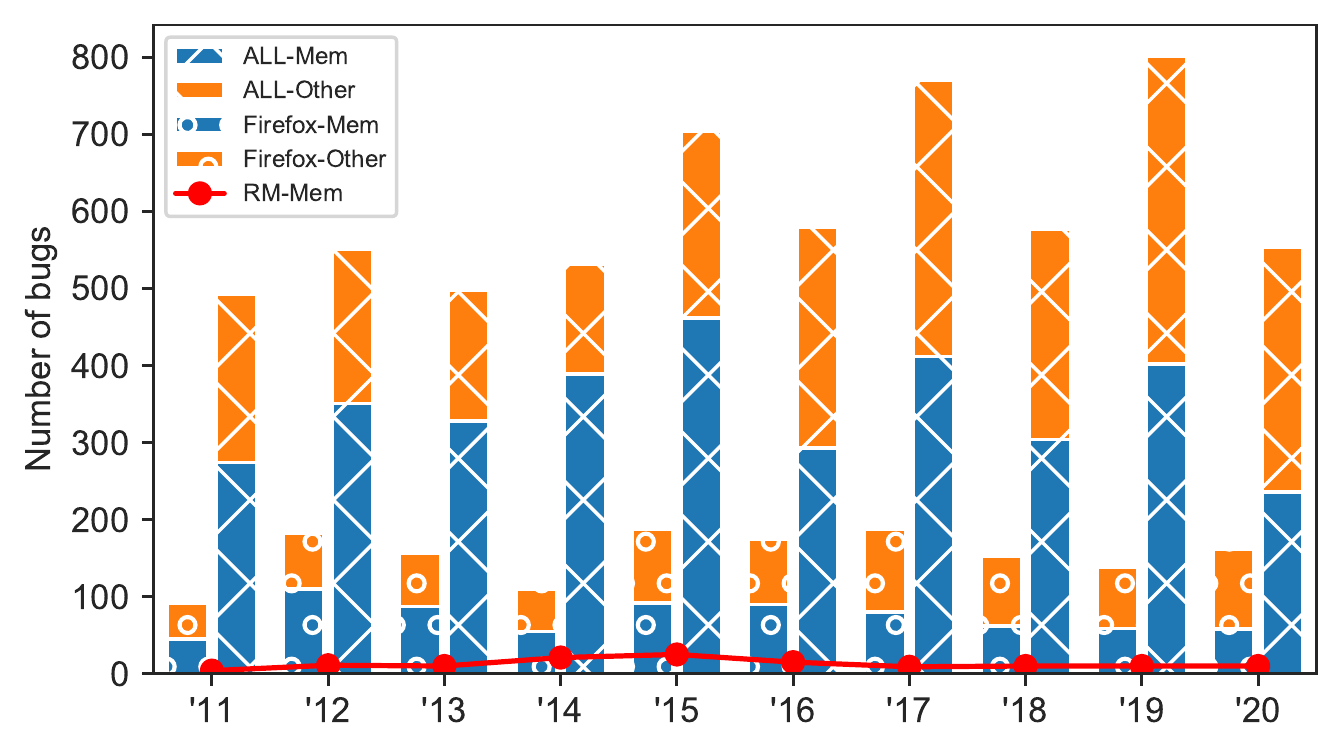}
        \caption{
        The number of memory-safety bugs vs. other bugs 
        in Firefox and other browsers.
        \textbf{RM-Mem} is the number of memory-safety bugs in the
        media parsing component in Firefox's renderer, 
        depicting a decline after it was partially
        rewritten in Rust starting in 2015.
        }\label{fig:mem-safety-bugs}
    \end{minipage}
    \hfill
    \begin{minipage}[t]{0.32\textwidth}
        \vspace{0pt}
        \centering
        \includegraphics[width=\columnwidth]{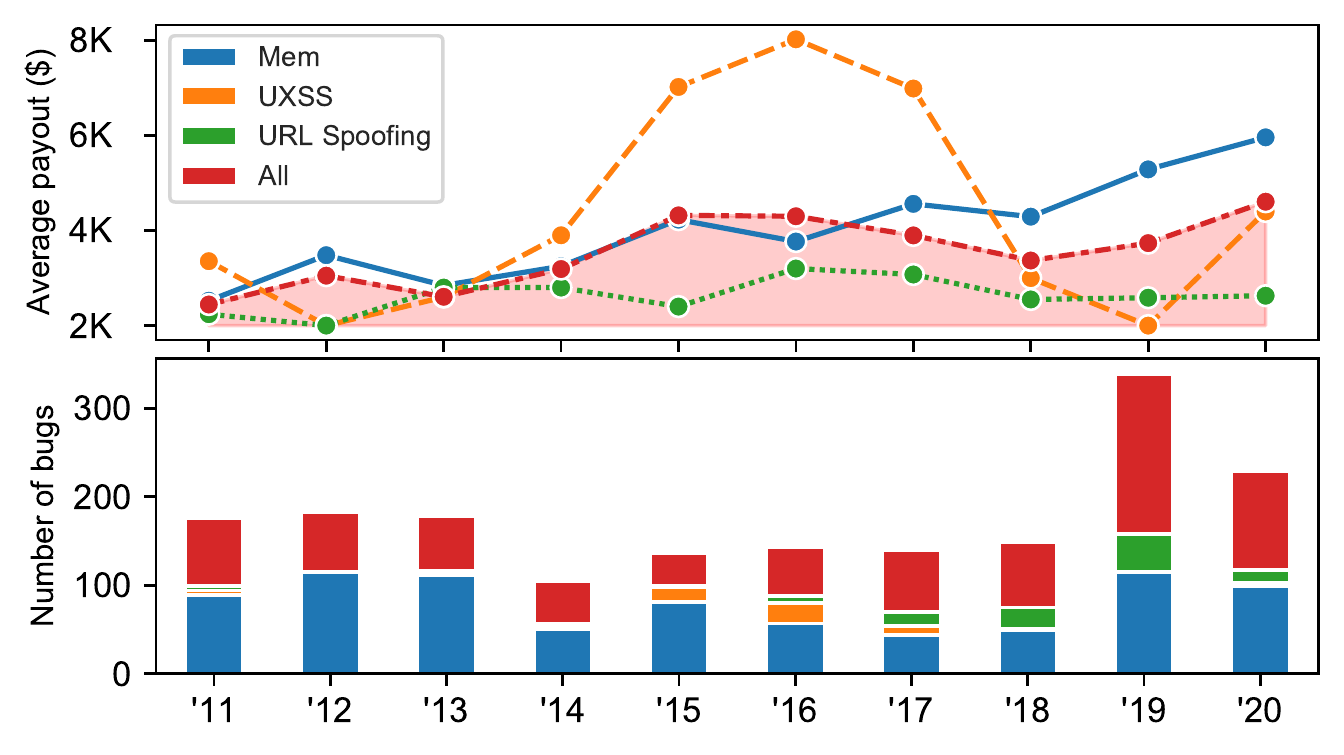}       
        \caption[bug bounty fig]{
        Correlations between average payout amounts (top chart) to the yearly
        number of bugs in Chromium (bottom chart). The red area is the average
        bounty amount for all classes\footnotemark. Bug classes crossing the red area
        indicate a higher bounty than the average.
        }\label{fig:bug-bounty}
    \end{minipage}
    \hfill
    \begin{minipage}[t]{0.32\textwidth}
        \vspace{0pt}
        \centering
        \includegraphics[width=\columnwidth]{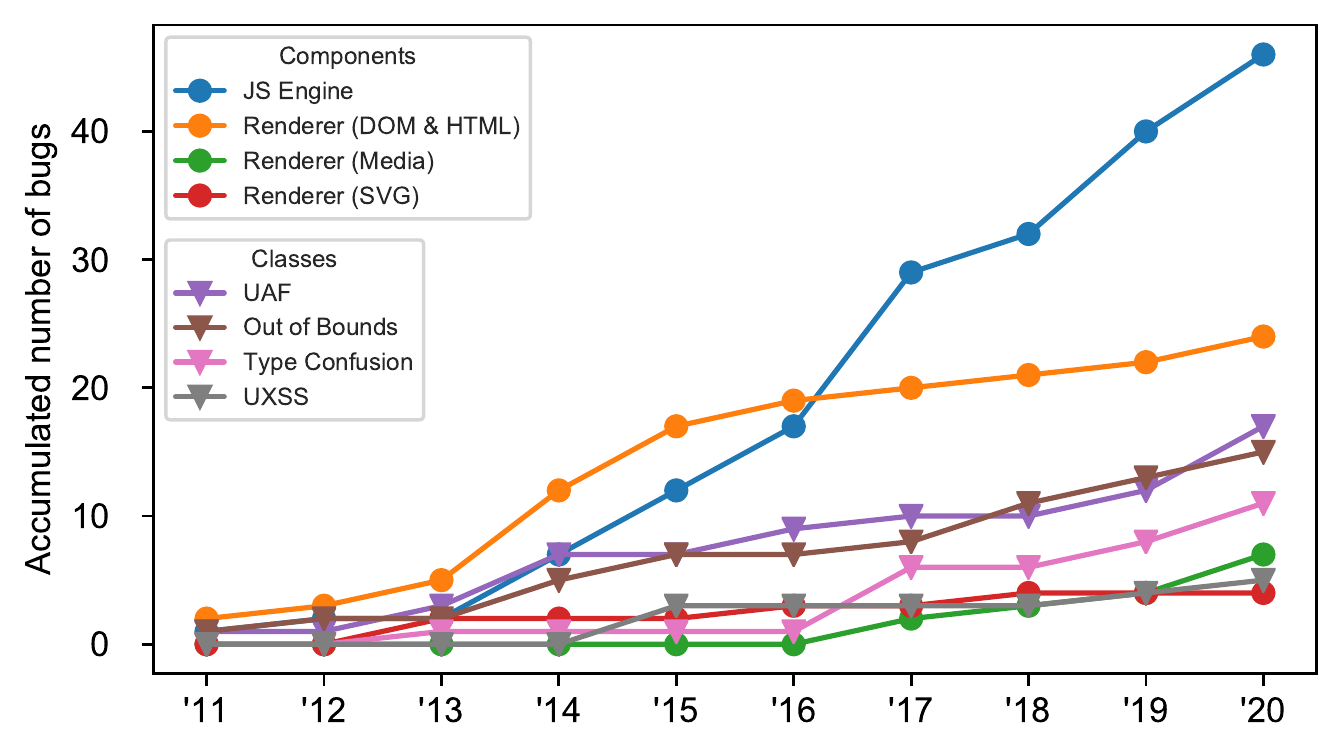}
        \caption{
        The trends of browser components and classes of exploited bugs. The data
        include bugs from all browsers. 
        Lines are the accumulated numbers. The
        \js engine and UAF bugs are dominating the exploited bug
        components and classes, respectively.
        }\label{fig:exp-bugs}
    \end{minipage}
\vspace{-13pt}
\end{figure*}

\PP{Memory-safe language.}
Memory-safety bugs are critical and dominant in browsers. For example, 
Chromium labels over 70\% of their high severity bugs as memory-safety issues, 
half of which are UAF bugs~\cite{chr-mem-bugs}. We show the ratio of 
memory-safety bugs in browsers in~\autoref{fig:mem-safety-bugs}. 
As shown in the figure, memory-safety bugs remain dominant for the past decade 
despite existing mitigations~\cite{google-safer-cpp}~\cite{ff-smt-ptr-gl}. 
Recently, there have been efforts in rewriting browsers 
using memory-safe languages 
(\eg, Rust) to mitigate memory-safety bugs.
For example, Mozilla is rewriting
parts of Firefox in Rust with an ongoing project
called Oxidation~\cite{oxid}.
Up until 2020, the Oxidation project had replaced 12\% of Firefox's
components with Rust equivalents. Five of the replaced subcomponents fall under 
the renderer's media parsing component.
We also plot the number of memory-safety bugs in the renderer's media parsing 
component in \autoref{fig:mem-safety-bugs}. It is clear that the number of 
memory-safety bugs in Firefox has shown 
a small but steady decline since Oxidation 
started in 2015, with a noticeable drop 
in memory-safety bugs in the renderer's 
media component. 
Despite several attempts from browser vendors to
counter memory safety issues, none of them resulted in a high impact like
Firefox's Oxidation.

\lesson{Using memory safe languages is 
an effective mitigation against memory-safety bugs}{
As shown in \autoref{fig:mem-safety-bugs},
the use of Rust in Firefox effectively reduces the memory safety bugs.
Though it takes a lot of effort, 
it is a fundamental way, 
and the most promising way 
to eliminate memory safety bugs.
We suggest that other browser vendors follow this best practice,
and gradually shift their browsers to memory-safe languages.
}

\PP{Bug bounty programs.}
Major browser vendors such as Google provide rewards
for proper security bug reports that help them to fix vulnerabilities
~\cite{chrome-bounty-program}.
In most cases, these payouts account for multiple factors such as
bug type, exploitability, and significant additional effort made by the reporter.
Higher payouts indicate higher incentives for researchers and whitehats to 
find bugs. We correlate the average payout amounts per year to the
yearly number of bugs in Chromium in~\autoref{fig:bug-bounty}. 
We particularly show memory-safety,
UXSS, and URL spoofing bugs since they depict interesting patterns with respect
to their payout amounts.

Payout amounts have an influence on the number of bugs 
found for respective classes (\autoref{fig:bug-bounty}).
Bug classes that have had average bounty amounts above the overall average
amount (\eg, UXSS in 2014-2016 and Mem bugs in 2017-2020) 
seem to
increase in numbers on those exact years.
This correlation does not work both ways, however: an increase of the number of
bugs of a certain class does not encourage higher bounty amounts. 
This figure further emphasizes other important factors that guide
researchers' efforts while looking for bugs besides seeking higher payouts, 
such as 
1) seeking the perks of exploring uncharted attack vectors (URL spoofing), 
2) aiming for bugs with higher impact (UXSS increased in 2016), 
and 3) avoiding bug classes that have effective mitigations (Site Isolation 
released in 2017 and UXSS bugs decreased in 2018).

\lesson{Higher payouts motivate more bug reports}{
    Browsers try to increase coverage and payout of bug bounty programs,
    which led to more bug reports.
    Therefore, increasing bug bounty payouts 
    can effectively attract the interest of security researchers
    and reduce attack surfaces.
}

\PP{Divergence of bug severity ratings.}
The Common Vulnerability Scoring System (CVSS)~\cite{first-cvss} was developed
as a free and open source standard for bug severity assessment.
The National Vulnerability Database~\cite{nvd} uses the CVSS standard to provide
bug severity base scores for each issued CVE number.
Similarly, Firefox and Chromium provide an assessment of a bug's severity in
their bug trackers~\cite{chr-bug-trac}~\cite{ff-bug-trac} and security
advisories~\cite{ff-sec-adv} but use their own bug rating systems.
\autoref{t:bug-ratings}~compares the bug severity assessments of NVD's CVSS-V3 
against those of Firefox and Chromium.
The aim of this study is to measure the effectiveness of using NVD's CVSS-V3
scores as a unified scale for bug severity in browsers.

In the table, we notice a divergence between the rating systems (Vendor vs. NVD).
NVD rates more than half of Firefox's bugs with
different severity scores, while it is only in agreement with Chromium
on 58\% of Chromium's bugs.
While full agreement between the two rating systems is not expected, the
big gap between the ratings is surprising. 
The divergence between ratings also occurs between vendors.
Looking at the number of critical bugs in the two browsers, 
we can see that Firefox assigns a much higher percentage of its bugs
as Critical compared to Chromium.
Our analysis results align with previous concerns regarding the use
 of CVSS scores as a metric for bug triaging and prioritizing
 ~\cite{tenable-stop-cvss}~\cite{cmu-improve-cvss}~\cite{mcafee-stop-cvss}.

\footnotetext{CVE-2011-3046~\cite{CVE-2011-3046}, the largest bounty rewarded by
Chromium (\$60k), is an outlier that was removed from the figure.}

\begin{table}[t]
    \centering
    \footnotesize
    \setlength{\tabcolsep}{3pt}
    \input{tbl/vendor-nvd-divergence}

    \caption{
        A comparison between NVD-assigned CVSS-V3 scores 
        vs. vendors' bug
        severity scores.  
        Bug ratings are: \textbf{L}ow, \textbf{M}oderate,
        \textbf{H}igh, and \textbf{C}ritical.  
        Bugs that are too
        old\protect\footnotemark~to have CVSS-V3 scores 
        are omitted from the
        table. The last column is the total 
        number of bugs used for this
        comparison, and the number of bugs that 
        have different bug ratings assigned.
        The table emphasizes the divergence of 
        bug ratings in two dimensions:
        1) CVSS scores vs. vendors' scores as seen 
        in the number of bugs triaged 
        as \textbf{L}ow (emboldened), and 
        2) bugs rated as \textbf{C}ritical 
        from vendor to vendor (underlined).
    }
    \label{t:bug-ratings}
    \vspace{-13pt}
\end{table}

\footnotetext{NVD started to use the CVSS-V3 scoring system in 2015
which has four levels of ratings, matching vendors'
rating levels, which justifies our comparison.} 

\PP{Bugs in browser exploits}
Bugs used in real-world browser exploits deserve more attention, as they are
indications of favorable attack vectors from the attackers' point of view.
To study such bugs, we collect information from credible sources
that only acquire highly exploitable bugs.
For bugs used in the wild, we refer to Google's routinely updated Project Zero
report, which tracks all publicly known cases of zero-day
exploits since 2014~\cite{p0-itw-bugs}.
We also collect the bugs exploited in Pwn2Own~\cite{pwn2own2020-rule}, 
a real-world hacking competition sponsored by the 
Zero Day Initiative~\cite{zdi}.
We highlight the top exploited bugs in the past decade in~\autoref{fig:exp-bugs},
based on their bug class and the host browser components.

As shown in~\autoref{fig:exp-bugs},
for browser components, 
DOM bugs were dominant until they were overtaken by JS engine bugs in 2017. 
Nevertheless, DOM bugs remain relevant today 
and show a slow but steady
increase even after adding many mitigations.
For bug type, memory-safety bugs such as UAF bugs are still favorable over 
other bug classes in real-world exploits 
despite all the mitigations in place.
One interesting observation is the trend of 
emerging bug classes and components.
For most of the lines in the figure, we see a rather steep increase
in the early years, but the increase slows down afterward 
(except for \js engine bugs).
This trend visualizes the attacker's endless effort to find and explore
new attack techniques, and the vendors' reactive 
countermeasures to eliminate 
and mitigate new bugs. \looseness=-1

%% file: tbl/vendor-nvd-divergence.tex
 
\begin{tabular}{c c c c c c c c c c}
        \toprule
        \multirow{2}{*}{Browser} & \multicolumn{4}{c}{Vendor} & \multicolumn{4}{c}{NVD's CVSS-V3} &
        \multirow{2}{*}{Total (Diff)} \\
        \cmidrule{2-9}

        & L & M & H & C & L & M & H & C & \\
        \midrule

        Firefox  & \textbf{13\%} & 34\% & 35\% & \underline{17\%} & \textbf{1\%} & 37\% & 39\% & \underline{22\%} & 947 (491) \\

        Chromium & \textbf{16\%} & 39\% & 43\% & \underline{2\%} & \textbf{1\%} & 46\% & 47\% & \underline{6\%} & 1191 (505) \\

        \bottomrule

\end{tabular}

%% file: sok-2.tex
\section{More Security Mitigations in Browsers}
\label{s:mitigation}
In this section, we present more generic mitigations implemented by 
browser vendors that are not covered in previous sections. 
We present a longitudinal study on the mitigations implemented in the 
four major browsers in the past decade, as well as the dates when they 
were applied and retired, in \autoref{t:mitigation}. In this section, 
we discuss a few of them in detail. 

\begin{table*}
    \centering
    \footnotesize
    \input{tbl/mitigation}
    \caption{Mitigations in browsers.}
    \vspace{-13pt}
    \label{t:mitigation}
\end{table*}

\subsection{Sandbox}
The sandbox is crucial to browser security because 
it confines the effect of bugs in
the renderer process, 
which contains various error-prone components. 
Except for cases like UXSS, attackers need to 
escape from the renderer sandbox 
using an exploit for the kernel, system services, 
or the browser process.
Consequently, it significantly raises the bar 
for attacks because attackers 
need to exploit both components 
(renderer and sandbox) to have a full-chain 
0-day exploit.

\PP{Win32k lockdown}
Since most Windows kernel vulnerabilities 
have been in Win32k system calls, 
Microsoft introduced the System Call Disable Policy
aka. Win32k lockdown~\cite{win32k-lockdown}
for Windows in 2012. 
This allows the developer of a Windows application 
to completely block access to the 
Win32k system call table, 
significantly reducing the attack surface.
Edge, Chrome, and Firefox have 
already adopted this mechanism to protect the browsers.
As a result, achieving sandbox escape from 
the renderer process has become much more complex.

\PP{Hypervisor based sandboxing}
Windows Defender Application Guard (WDAG)~\cite{WDAG} was introduced by 
Microsoft to isolate untrusted websites 
or resources (\eg, files) in 
enterprise scenarios.
WDAG uses Hyper-V to create a 
fresh instance of Windows at the hardware layer, 
which includes a separate copy of the kernel
and the minimum Windows Platform Services
to make sure that the Edge browser functions
normally.
WDAG is implemented in Edge to 
protect against advanced attacks that can bypass 
the browser sandbox. 
With WDAG, the attacker needs to escape 
both the sandbox and the 
Hyper-V virtual machine.

\subsection{Hardened Allocators}
\label{s:hardened-allocators}
Browsers use specialized heap allocators for many objects for performance 
and security reasons~\cite{oilpan, riptide-gc}.
These allocators use specific designs 
which help reduce damage by limiting attack primitives.

\PP{Isolated heap} 
Isolated heap is an effective defense to prevent use-after-free (UAF) 
attacks being escalated to type confusion attacks.
By isolating objects based on their 
1) type, 
2) security hazard level (\eg, embedding v-table pointer), and 
3) \js reachability (\eg, \cc{ArrayBuffer}), 
isolated heap 
effectively raises the bar for UAF exploitation.
The isolation prevents an attacker from re-claiming the freed object with 
an object with a different layout, which is typical for exploiting UAFs 
in browsers.

Modern browsers implement a basic level of heap separation 
between \js-reachable objects and other 
objects~\cite{ff-isoheap,fsecure-gigacage,hariri2015abusing,partitionalloc}. 
However, it was still possible to create type confusion via UAF among 
the objects in the same heap but in other types.
To prevent this attack, 
Safari~\cite{baumstarkniklas19masterarbeit} 
and Firefox~\cite{ff-presarena} 
introduced separate heaps for every
type in specific categories, 
which provided a much more 
fine-grained isolation. 
Therefore, there is no public, 
generic exploitation methodology 
for exploiting UAF bugs in all browsers.

\PP{Delayed free.}
%UAF
Another mitigation, \emph{elayed free}, 
effectively increases the difficulty 
of exploiting UAF bugs, 
but this approach cannot 
restrict the reclamation of dangling 
pointers.
Browsers use various garbage collection (GC) algorithms
to deallocate heap-allocated objects with no references.
Some variants of GC additionally scan
stack and heap~\cite{oilpan, riptide-gc} areas
to find possibly overlooked references, which is known as
\emph{conservative scanning}~\cite{conservative-gc}
or \emph{delayed free}~\cite{hariri2015abusing}.
Notably, \ff dropped this in favor of \emph{exact rooting}
and wrote a static analysis tool to 
find unsafe usage of references from the stack
\cite{ff-stack-rooting,ff-root-analysis}.
\chrome also has a similar tool \cite{gcmole}, 
but it is only enforced
on specific areas.
However, \emph{delayed free} has 
introduced side-channel primitives
that can be used to defeat ASLR since it cannot
distinguish pointers to the heap 
and user-controlled integers by 
design~\cite{hariri2015abusing, dedup, dude, gc2021sidechannel}.

\PP{Heap metadata protection.}
Heap metadata protection is an approach that checks the metadata portion 
of heap chunks to prevent  heap corruption and silent error propagation    
in the heap.
For example, a
heap allocator may
put a random value~\cite{vista-heap}
before dangerous data structures
to detect
heap exploits.
\emph{PartitionAlloc} in Chrome removed in-line metadata and placed 
guard pages
to prevent linear heap overflow from 
overwriting metadata~\cite{partitionalloc}.
There are also some OS-level efforts on metadata 
protection~\cite{vista-heap,scudo-allocator}.

\PP{Other mitigations on heap}
\emph{Frame poisoning} in \ff deallocated chunks of memory 
with addresses pointing to non-accessible memory pages~\cite{ff-presarena}.
Similarly, in Edge, this is done by filling zeros when freeing 
heap chunks~\cite{yason2015understanding}.
\emph{GWP-ASan}~\cite{gwp-asan} in Chrome randomly places a small portion 
of allocated objects right before/after guard pages
and deallocates the entire page when the chunk is freed to detect 
heap errors in the wild.

\subsection{Control-Flow Integrity}
Since attackers often manipulate 
the values of instruction pointers to achieve 
code execution, control-flow integrity 
is enforced to prevent them from 
hijacking control flows, making the attack more difficult.
The compiler infrastructure, OS and hardware support 
provide most mitigations in this category,  
such as protecting virtual function tables by
introducing canary values~\cite{vtguard} 
and allowing listing indirect branches
by checking the destination address~\cite{clang-cfi,cfg-msvc}. 

There is ongoing work
to prevent arbitrary memory writes from modifying code regions 
that are executable by attackers (\autoref{s:overwrite-protection}).
Based on hardware support, browsers could apply additional mitigations 
without a dramatic decrease of performance, such as adding pointer integrity 
checks using PAC on ARM64~\cite{pac} and adding additional 
W $\oplus$ X protection on JIT-compiled code using Intel MPK~\cite{pku-wasm} 
and APRR~\cite{siguza-appr}.

\subsection{Side Channels}
\label{s:mitigation:sc}
Browsers are also vulnerable to side-channel attacks. To date, studies have 
shown that sensitive information in browsers can be inferred via 
1) microarchitectual state~\cite{shusterman2021prime,lipp2018meltdown,
kocher2019spectre,mcilroy2019spectre,oren2015spy}; 
2) GPU\cite{lee2014stealing,naghibijouybari2018rendered}; 
3) floating-point timing channels~\cite{andrysco2015subnormal} and
4) browser-specific side channels~\cite{vila2017loophole,smith2018browser,
van2015clock,van2020timeless}.
Researchers have introduced defense mechanisms
\cite{schwarz2018javascript, snyder2017most,
cao2017deterministic,kohlbrenner2016trusted} to 
protect the browsers from side-channel attacks, such as 
DeterFox~\cite{cao2017deterministic}
and FuzzyFox~\cite{kohlbrenner2016trusted}.
Also, browser vendors have implemented 
defenses~\cite{chrome-mitigate-sidechannel,mozilla-mitigate-sidechannel,
safari-spectre}, 
which can be classified into two categories as follows:

\PP{Reduce the resolution of timers.}
Since most of the attacks rely on accurate timing, 
to hinder the detection of small timing differences, browser vendors reduced 
the resolution of the precise timer (\eg, \cc{performance.now()}) and 
introduced random jitters to prevent resolution 
recovery~\cite{chrome-timer-1,webkit-timer-1,firefox-timer-1}. After the 
discovery of Spectre~\cite{kocher2019spectre} and 
Meltdown~\cite{lipp2018meltdown}, the vendors further lowered the precisions 
of timers~\cite{chrome-timer-spectre,ie-timer-spectre,ff-timer-spectre}.
Since \cc{SharedArrayBuffer} can be used to create a high-resolution timer, 
shortly after the discovery of Spectre~\cite{kocher2019spectre}, 
\cc{SharedArrayBuffer} was disabled in all modern 
browsers~\cite{spectre-mitigation-news1}.

\PP{Prevent resource sharing.}
Another mitigation technique is to 
prevent resource sharing between the victim and 
the attacker. Site 
Isolation~\cite{reis2019site,siteisol} (\autoref{s:arch:diff}) 
effectively mitigates the Javascript-based transient execution.
Cross-Origin-Opener-Policy (COOP) and Cross-Origin-Embedder-Policy 
(COEP)~\cite{coop-coep-explain} were introduced to set up a cross-origin 
isolated environment. COOP allows a website to include a response header on
a top-level document, ensuring that the cross-origin documents do not
share the same browsing context group with itself, thus preventing direct DOM 
access. COEP prevents a page from loading any cross-origin resources that 
do not give explicit permission to this page. These are both enforced 
using HTTP headers, and they were shipped in Chrome 83, Firefox 79, and
Edge 83~\cite{coop-MDN,coep-MDN}, 
while \safari does not support 
them yet as of November 2021.

After the introduction of Site Isolation and COOP/COEP, Chrome and Firefox 
re-enabled the use of \cc{SharedArrayBuffer}~\cite{chrome-sab-reenable,
ff-sab-reenable}. However, a study showed that the re-introduction of 
\cc{SharedArrayBuffer} increases the covert channel 
capacity 2,000-fold and 800,000-fold, respectively~\cite{rokickisok}. 
Two recent papers~\cite{spookjs,jin2022timing} suggested that 
despite Site Isolation in Chrome, 
the attacker can still learn cross-origin sensitive information.

\subsection{Other Mitigation Efforts}
\PP{UAF mitigation.}
To fundamentally fix the UAF problems that are not covered by 
garbage collection or other safety measures, the Chrome team introduced a 
term called \textit{MiraclePtr}~\cite{miracleptr1,miracleptr2}, which stands 
for a set of algorithms that can wrap raw pointers in C/C++ so that they 
are not exploitable via UAF. 
\textit{MiraclePtr} is expected to be in production soon~\cite{miracleptr3}.

\PP{Improving memory safety.}
The Chrome team has explored improvements 
for their C++ codebase that can eliminate/reduce 
specific types of bugs by limiting the use of specific language features 
(\eg, C++ exceptions~\cite{google-cpp-guidelines}) and introducing wrapper 
classes around integer operations~\cite{google-safer-cpp}.

\PP{Improving JIT compiler.}
There have been efforts to safeguard dangerous optimizations inside 
JIT compilers.
For example, many exploits make use of bounds check elimination~\cite{saelo18}
that removes seemingly redundant bounds checks. 
To mitigate this, the Chrome team introduced a patch that marked such checks 
as \textit{aborting} instead of simply removing them~\cite{abort-bound-check}. 
Therefore, the attacker can only trigger a \cc{SIGTRAP} at best.
Moreover,
to make bytecode generation for standard JS functions less error-prone, 
the Chrome team made a domain-specific language,
\emph{Torque}~\cite{torque}, 
which replaced the 
existing C++ implementations and reduced a lot of LoC. 

\lesson{Collaborative efforts on mitigations are good}
{When one vendor deploys a mitigation, other vendors are likely to follow. 
In~\autoref{t:mitigation}, we saw that most of 
the mitigations have been adopted 
by multiple browsers. 
If there are bugs found in one browser, 
the vendor can quickly share the 
information with other vendors and they can work 
together to build better mitigations using collective knowledge.
In the case of Spectre/Meltdown attacks~\cite{kocher2019spectre,
lipp2018meltdown}, browser vendors worked together to build a plan for 
mitigating the immediate threats~\cite{chrome-timer-spectre,ie-timer-spectre,
ff-timer-spectre}, which is a great example of collaborative effort.}

%% file: tbl/mitigation.tex
% Per-browser symbols
\newcommand{\Chrome}{$\dagger$}
\newcommand{\Edge}{$\ddagger$}
\newcommand{\Firefox}{$\sharp$}
\newcommand{\Safari}{$\flat$}
\setlength{\tabcolsep}{0.2pt}
\raggedleft\start: Mitigation applied, \no: Mitigation retired \\
\textbf{\red{C}:} Chrome, \textbf{F:} Firefox, \textbf{\green{S}:} Safari, \textbf{\blue{I/E}:} IE / Edge
\begin{tabular}{@{}|c|l||c|c|c|c||cccc|cccc|cccc|cccc|cccc|cccc|cccc|cccc|cccc|cccc|cccc|@{}}

\hline
\multicolumn{2}{|c||}{\multirow{2}{*}{Attack}}
  & \multicolumn{4}{c}{}
  & \multicolumn{16}{l}{$\downarrow$Bypass OS mitigations\cite{CVE-2003-1048,CVE-2006-5579}}
  & \multicolumn{2}{c}{}
  & \multicolumn{10}{l}{$\downarrow$Bypass CFG\cite{yunhai2015bypass}}
  & \multicolumn{4}{c}{$\downarrow$JS\cite{jit-exploitation}}
  & \multicolumn{4}{c}{}
  & \multicolumn{6}{l}{$\downarrow$Glitch\cite{frigo2018grand}}
  & \multicolumn{2}{c|}{} \\

\multicolumn{2}{|l||}{}
  & \multicolumn{4}{c}{}
  & \multicolumn{12}{l}{$\downarrow$Bypass SELinux\cite{selinux-leak-fd}}
  & \multicolumn{14}{l}{$\downarrow$Sandbox escape\cite{CVE-2013-6632}}
  & \multicolumn{6}{c}{$\downarrow$JS core\cite{CVE-2016-4622}}
  & \multicolumn{4}{l}{$\downarrow$S/M\cite{kocher2019spectre}}
  & \multicolumn{8}{l|}{$\downarrow$Code sign\cite{code-sign-bypass}}  \\
\hline

\multicolumn{2}{|c||}{\multirow{2}{*}{Mitigation}}
  & \multicolumn{4}{c||}{Browser}
  & \multicolumn{4}{c|}{2010}
  & \multicolumn{4}{c|}{2011}
  & \multicolumn{4}{c|}{2012}
  & \multicolumn{4}{c|}{2013}
  & \multicolumn{4}{c|}{2014}
  & \multicolumn{4}{c|}{2015}
  & \multicolumn{4}{c|}{2016}
  & \multicolumn{4}{c|}{2017}
  & \multicolumn{4}{c|}{2018}
  & \multicolumn{4}{c|}{2019}
  & \multicolumn{4}{c|}{2020} \\

\cline{3-50}

  \multicolumn{2}{|l||}{}
  & \red{C} & F & \green{S} & \blue{I/E}
  & 1 & 2 & 3 & 4
  & 1 & 2 & 3 & 4
  & 1 & 2 & 3 & 4
  & 1 & 2 & 3 & 4
  & 1 & 2 & 3 & 4
  & 1 & 2 & 3 & 4
  & 1 & 2 & 3 & 4
  & 1 & 2 & 3 & 4
  & 1 & 2 & 3 & 4
  & 1 & 2 & 3 & 4
  & 1 & 2 & 3 & 4 \\ \hline\hline

\multirow{3}{*}{JIT}
  & StructureID randomization~\cite{webkit-structureid}
  &   &   & \yes  &
  &   &   &   &
  &   &   &   &
  &   &   &   &
  &   &   &   &
  &   &   &   &
  &   &   &   &
  &   &   &   &
  &   &   &   &
  &   &   &   &
  &   &   &   & \G\green\start
  & \G  & \G  & \G  & \G \\ \cline{2-50}

  & GigaCage~\cite{fsecure-gigacage}
  &   &   & \yes  &
  &   &   &   &
  &   &   &   &
  &   &   &   &
  &   &   &   &
  &   &   &   &
  &   &   &   &
  &   &   &   &
  &   &   & \G\green\start  & \G
  & \G  & \G  & \G  & \G
  & \G  & \G  & \G  & \G
  & \G  & \G  & \G  & \G \\ \cline{2-50}

  & PACCage~\cite{pzero-jit}
  &   &   & \yes  &
  &   &   &   &
  &   &   &   &
  &   &   &   &
  &   &   &   &
  &   &   &   &
  &   &   &   &
  &   &   &   &
  &   &   &   &
  &   &   &   &
  &   & \G\green\start  & \G  & \G
  & \G  & \G  & \G  & \G \\ \hline

  % & JIT Hardened Range Check
  % & \yes  &   &   &
  % &   &   &   &
  % &   &   &   &
  % &   &   &   &
  % &   &   &   &
  % &   &   &   &
  % &   &   &   &
  % &   &   &   &
  % &   &   &   &
  % &   &   &   &
  % &   &   &   & \G\red\start
  % & \G  & \G  & \G  & \G \\ \cline{2-50}

  % & Controlled bytes elimination~\cite{gawlik2018sok}
  % &   &   & \yes  & \yes
  % &   &   &   &
  % &   &   &   &
  % & \G\green\start  & \G  & \G  & \G
  % & \G  & \G  & \G  & \G
  % & \G  & \G  & \G  & \G
  % & \G  & \G  & \G\blue\start  & \G
  % & \G  & \G  & \G  & \G
  % & \G  & \G  & \G  & \G
  % & \G  & \G  & \G  & \G
  % & \G  & \G  & \G  & \G
  % & \G  & \G  & \G  & \G \\ \cline{2-50}

  % & Internal Randomization~\cite{gawlik2018sok}
  % &   &   & \yes  & \yes
  % &   &   &   &
  % &   &   &   &
  % & \G\green\start  & \G  & \G  & \G
  % & \G  & \G  & \G  & \G
  % & \G  & \G  & \G  & \G
  % & \G  & \G  & \G\blue\start  & \G
  % & \G  & \G  & \G  & \G
  % & \G  & \G  & \G  & \G
  % & \G  & \G  & \G  & \G
  % & \G  & \G  & \G  & \G
  % & \G  & \G  & \G  & \G \\ \hline

  \multirow{5}{*}{W$\oplus$X}
  & W$\oplus$X protection~\cite{firefox-rwx}
  & \yes  & \yes  & \yes  & \yes
  &   &   &   &
  &   &   & \G\red\start  & \G
  & \G\green\start  & \G  & \G  & \G
  & \G  & \G  & \G  & \G
  & \G\start  & \G  & \G  & \G
  & \G  & \G  & \G\blue\start  & \G
  & \G  & \G  & \G  & \G
  & \G  & \G  & \G  & \G
  & \G  & \G  & \G  & \G
  & \G  & \G  & \G  & \G
  & \G  & \G  & \G  & \G \\ \cline{2-50}

& Hardened runtime~\cite{safari-hardened-runtime}
  & \yes  & \yes  & \yes  &
  &   &   &   &
  &   &   &   &
  &   &   &   &
  &   &   &   &
  &   &   &   &
  &   &   &   &
  &   &   &   &
  &   &   & \G\red\start  & \G
  & \G  & \G  & \G\start  & \G
  & \G  & \G  & \G\green\start  & \G
  & \G  & \G  & \G  & \G \\ \cline{2-50}

  & Out-of-process JIT~\cite{edge-acg}
  &   &   &   & \yes
  &   &   &   &
  &   &   &   &
  &   &   &   &
  &   &   &   &
  &   &   &   &
  &   &   &   &
  &   &   &   &
  & \G\blue\start  & \G  & \G  & \G
  & \G  & \G  & \G  & \G
  & \G  & \G  & \G  & \G
  & \G  & \G  & \G  & \G \\ \cline{2-50}

  % While chrome-signed (crbug/750886) suggests CIG on Windows, it's actually applied to renderers in 2017.08
  & Code signing enforcement~\cite{edge-acg,chrome-signed,firefox-signed}
  & \yes  & \yes  &   & \yes
  &   &   &   &
  &   &   &   &
  &   &   &   &
  &   &   &   &
  &   &   &   &
  &   &   &   & 
  &   &   &   & 
  & \G\blue\start  & \G  & \G  & \G
  & \G  & \G  & \G  & \G
  & \G\red\start  & \G  & \G  & \G\start
  & \G  & \G  & \G  & \G \\ \cline{2-50}

  & Arbitrary code guard~\cite{edge-acg,chrome-acg,firefox-acg}
  & \yes  & \yes  &   & \yes
  &   &   &   &
  &   &   &   &
  &   &   &   &
  &   &   &   &
  &   &   &   &
  &   &   &   &
  &   &   &   &
  & \G\blue\start  & \G  & \G  & \G
  & \G  & \G  & \G  & \G
  & \G\start  & \G  & \G\red\start  & \G
  & \G  & \G  & \G  & \G \\ \hline

  \multirow{2}{*}{Heap}
  & Isolated heap~\cite{yason2015understanding,fsecure-gigacage,partitionalloc,ff-isoheap}
  & \yes & \yes  & \yes & \yes
  &   &   &   &
  &   &   &   &
  &   &   &   &
  &   & \G\red\start  & \G  & \G
  & \G  & \G\blue\start  & \G  & \G
  & \G  & \G  & \G  & \G
  & \G  & \G  & \G  & \G
  & \G  & \G  & \G  & \G \start
  & \G\green\start  & \G  & \G  & \G
  & \G  & \G  & \G  & \G
  & \G  & \G  & \G  & \G \\ \cline{2-50}

  & Delayed free~\cite{yason2015understanding, riptide-gc, ff-stack-rooting}
  & \yes  & \yes  & \yes  & \yes
  &   & \G\start  & \G  &\G
  & \G  & \G  & \G  &\G
  & \G  & \G  & \G  &\G
  & \G  & \G  & \G  &\G
  & \G\no  &   & \G\blue\start  & \G
  & \G  & \G  & \G  & \G\red\start
  & \G  & \G  & \G  & \G\green\start
  & \G  & \G  & \G  & \G
  & \G  & \G  & \G  & \G
  & \G  & \G  & \G  & \G
  & \G  & \G  & \G  & \G \\ \hline

% Branchless Security Checks (Safari, Chrome) mean:
% - Index Masking (Later deprecated / replaced with poisoning)
% - Pointer Poisoning
% These are now all removed. I've merged these, since it's incremental.
% \XXX{let's mention it in H/W mitigations}

\multirow{3}{*}{S/M}
  & Branchless security checks \cite{safari-spectre}
  &  \yes &   & \yes  & \yes
  &   &   &   &
  &   &   &   &
  &   &   &   &
  &   &   &   &
  &   &   &   &
  &   &   &   &
  &   &   &   &
  &   &   &   &
  & \G\green\start\kern-0.5em\blue\start  & \G\red\start  & \G\red\no  & \G
  & \G\green\no  & \G & \G & \G
  & \G & \G & \G & \G \\ \cline{2-50}

  & Disabling SharedArrayBuffer \cite{safari-spectre}
  & \yes & \yes  & \yes  & \yes
  &   &   &   &
  &   &   &   &
  &   &   &   &
  &   &   &   &
  &   &   &   &
  &   &   &   &
  &   &   &   &
  &   &   &   & \G\start
  & \multicolumn{2}{c}{\G\red\start\kern-0.5em\green\start\kern-0.5em\blue\start}  & \G\red\no  & \G
  & \G  & \G  & \G  & \G
  & \G  & \G  & \G\no  & \G \\ \cline{2-50}

  & Reducing timer accuracy~\cite{chrome-timer-spectre,ie-timer-spectre,ff-timer-spectre}
  & \yes  & \yes  & \yes  & \yes
  &   &   &   &
  &   &   &   &
  &   &   &   &
  &   &   &   &
  &   &   &   &
  &   &   &   &
  &   &   &   &
  &   &   &   &
  & \multicolumn{2}{c}{\G\red\start\kern-0.5em\start\kern-0.5em\green\start\kern-0.5em\blue\start}  & \G  & \G
  & \G  & \G  & \G  & \G
  & \G  & \G  & \G  & \G \\ \hline

  \multirow{2}{*}{CFI}
  & VTGuard~\cite{yason2015understanding}
  &   &   &   & \yes
  &   &   &   &
  &   &   &   &
  &   &   & \G\blue\start  & \G
  & \G  & \G  & \G  & \G
  & \G  & \G  & \G  & \G
  & \G  & \G  & \G  & \G
  & \G  & \G  & \G  & \G
  & \G  & \G  & \G  & \G
  & \G  & \G  & \G  & \G
  & \G  & \G  & \G  & \G
  & \G  & \G  & \G  & \G \\ \cline{2-50}

% TODO: While CFG is enabled on chrome, it does not add CFG instrumentation on its binaries
% CFG of windows protects other dlls as well, so it's turned on.
  & Control flow integrity~\cite{clang-cfi,cfg-msvc}
  & \yes  & \yes  &   & \yes
  &   &   &   &
  &   &   &   &
  &   &   &   &
  &   &   &   &
  &   &   &   & \G\blue\start
  & \G  & \G  & \G  & \G
  & \G  & \G  & \G  & \G
  & \G\red\start  & \G  & \G  & \G
  & \G\start  & \G  & \G  & \G
  & \G  & \G  & \G  & \G
  & \G  & \G  & \G  & \G \\ \hline

\multirow{9}{*}{SBX}
  & Seccomp (+bpf) sandbox~\cite{seccomp}
  & \yes  & \yes  &   &
  & \G\red\start  & \G  & \G  & \G
  & \G  & \G  & \G  & \G
  & \G  & \G  & \G  & \G
  & \G  & \G  & \G  & \G\start
  & \G  & \G  & \G  & \G
  & \G  & \G  & \G  & \G
  & \G  & \G  & \G  & \G
  & \G  & \G  & \G  & \G
  & \G  & \G  & \G  & \G
  & \G  & \G  & \G  & \G
  & \G  & \G  & \G  & \G \\ \cline{2-50}

  & Namespace (setuid, userns)~\cite{namespaces-linux}
  & \yes  & \yes  &   &
  & \G\red\start  & \G  & \G  & \G
  & \G  & \G  & \G  & \G
  & \G  & \G  & \G  & \G
  & \G  & \G  & \G  & \G
  & \G  & \G  & \G  & \G
  & \G  & \G\start  & \G  & \G
  & \G  & \G  & \G  & \G
  & \G  & \G  & \G  & \G
  & \G  & \G  & \G  & \G
  & \G  & \G  & \G  & \G
  & \G  & \G  & \G  & \G \\ \cline{2-50}

  & SELinux~\cite{selinux}
  & \yes  &   &   &
  &   & \G\red\start  & \G  & \G
  & \G  & \G  & \G  & \G
  & \G  & \G  & \G  & \G
  & \G  & \G\red\no  &   &
  &   &   &   &
  &   &   &   &
  &   &   &   &
  &   &   &   &
  &   &   &   &
  &   &   &   &
  &   &   &   & \\ \cline{2-50}

  & macOS sandbox~\cite{seatbelt}
  & \yes  & \yes  & \yes  &
  & \G\green\start  & \G  & \G  & \G
  & \G  & \G  & \G  & \G
  & \G  & \G  & \G  & \G
  & \G  & \G  & \G  & \G
  & \G\red\start\kern-0.5em\start  & \G  & \G  & \G
  & \G  & \G  & \G  & \G
  & \G  & \G  & \G  & \G
  & \G  & \G  & \G  & \G
  & \G  & \G  & \G  & \G
  & \G  & \G  & \G  & \G
  & \G  & \G  & \G  & \G \\ \cline{2-50}

  & Restricted tokens~\cite{restricted-token}
  & \yes  & \yes  &   & \yes
  & \G\red\start\kern-0.5em\blue\start  & \G  & \G  & \G
  & \G  & \G  & \G  & \G
  & \G  & \G  & \G  & \G
  & \G  & \G  & \G  & \G
  & \G  & \G\start  & \G  & \G
  & \G  & \G  & \G  & \G
  & \G  & \G  & \G  & \G
  & \G  & \G  & \G  & \G
  & \G  & \G  & \G  & \G
  & \G  & \G  & \G  & \G
  & \G  & \G  & \G  & \G \\ \cline{2-50}

  & AppContainer~\cite{yason2015understanding}
  &   &   &   & \yes
  &   &   &   &
  &   &   &   &
  &   &   &   & \G\blue\start
  & \G  & \G  & \G  & \G
  & \G  & \G  & \G  & \G
  & \G  & \G  & \G  & \G
  & \G  & \G  & \G  & \G
  & \G  & \G  & \G  & \G
  & \G  & \G  & \G  & \G
  & \G  & \G  & \G  & \G
  & \G  & \G  & \G  & \G \\ \cline{2-50}

  & Non-system font filtering~\cite{nonsystem-font}
  & \yes  &   &   &
  &   &   &   &
  &   &   &   &
  &   &   &   &
  &   &   &   &
  &   &   &   &
  &   &   &   &
  & \G\red\start  & \G  & \G  & \G
  & \G  & \G  & \G  & \G
  & \G  & \G  & \G  & \G
  & \G  & \G  & \G  & \G
  & \G  & \G  & \G  & \G \\ \cline{2-50}

  & Win32k lockdown~\cite{win32k-lockdown}
  & \yes  &   &   &
  &   &   &   &
  &   &   &   &
  &   &   &   &
  &   &   &   &
  &   &   &   &
  &   &   &   &
  &   &   &   &
  &   &   &   & \G\red\start
  & \G  & \G  & \G  & \G
  & \G  & \G  & \G  & \G
  & \G  & \G  & \G  & \G \\ \cline{2-50}

  & Plugin sandboxing (\eg, Flash)~\cite{sabanal2012digging}
  & \yes  & \yes  & \yes  & \yes
  &   &   &   &
  & \G\red\start  & \G  & \G  & \G
  & \G  & \G\start  & \G  & \G\blue\start
  & \G  & \G  & \G  & \G\green\start
  & \G  & \G  & \G  & \G
  & \G  & \G  & \G  & \G
  & \G  & \G  & \G  & \G
  & \G  & \G  & \G  & \G
  & \G  & \G  & \G  & \G
  & \G  & \G  & \G  & \G
  & \G  & \G  & \G  & \G \\ \hline

  % & Win32k Lockdown on Flash
  % & \yes  &   &   &
  % &   &   &   &
  % &   &   &   &
  % &   &   &   &
  % &   &   &   &
  % &   &   &   &
  % &   &   &   &
  % &   & \G\red\start  & \G  & \G
  % & \G  & \G  & \G  & \G
  % & \G  & \G  & \G  & \G
  % & \G  & \G  & \G  & \G
  % & \G  & \G  & \G  & \G \\ \cline{2-50}

  % & Extension Points Disabling
  % & \yes  &   &   &
  % &   &   &   &
  % &   &   &   &
  % &   &   &   &
  % &   &   &   &
  % &   &   &   &
  % &   &   &   &
  % &   & \G\red\start  & \G  & \G
  % & \G  & \G  & \G  & \G
  % & \G  & \G  & \G  & \G
  % & \G  & \G  & \G  & \G
  % & \G  & \G  & \G  & \G \\ \cline{2-50}

  \multirow{5}{*}{Other}
  & XSS auditor~\cite{xss-auditor}
  & \yes  &   & \yes  & \yes
  & \G\red\start\kern-0.5em\blue\start\kern-0.5em\green\start  & \G  & \G  & \G
  & \G  & \G  & \G  & \G
  & \G  & \G  & \G  & \G
  & \G  & \G  & \G  & \G
  & \G  & \G  & \G  & \G
  & \G  & \G  & \G  & \G
  & \G  & \G  & \G  & \G
  & \G  & \G  & \G  & \G
  & \G  & \G  & \G  & \G
  & \G  & \G  & \G  & \G\red\no
  & \G  & \G  & \G  & \G \\ \cline{2-50}

  & 64-bit ASLR
  & \yes  & \yes  & \yes  & \yes
  &   &   &   &
  &   &   & \G\green\start  & \G
  & \G  & \G  & \G  & \G\blue\start
  & \G  & \G  & \G\red\start  & \G
  & \G  & \G  & \G  & \G\start
  & \G  & \G  & \G  & \G
  & \G  & \G  & \G  & \G
  & \G  & \G  & \G  & \G
  & \G  & \G  & \G  & \G
  & \G  & \G  & \G  & \G
  & \G  & \G  & \G  & \G \\ \cline{2-50}

  & Hypervisor based sandboxing \cite{WDAG-edge1,WDAG-edge2}
  &   &   &   & \yes
  &   &   &   &
  &   &   &   &
  &   &   &   &
  &   &   &   &
  &   &   &   &
  &   &   &   &
  &   &   & \G\blue\start  & \G
  & \G  & \G  & \G  & \G
  & \G  & \G  & \G  & \G
  & \G  & \G  & \G  & \G
  & \G  & \G  & \G  & \G \\ \cline{2-50}

  & Site isolation \cite{reis2019site}
  & \yes  &   &   &
  &   &   &   &
  &   &   &   &
  &   &   &   &
  &   &   &   &
  &   &   &   &
  &   &   &   &
  &   &   &   &
  &   &   &   &
  &   & \G\red\start  & \G  & \G
  & \G  & \G  & \G  & \G
  & \G  & \G  & \G  & \G \\ \cline{2-50}

  & Pointer authentication code~\cite{pac}
  & \yes  &   & \yes  &
  &   &   &   &
  &   &   &   &
  &   &   &   &
  &   &   &   &
  &   &   &   &
  &   &   &   &
  &   &   &   &
  &   &   &   &
  &   &   & \G\red\start  & \G\green\start
  & \G  & \G  & \G  & \G
  & \G  & \G  & \G  & \G \\ \hline

\end{tabular}

%% file: pwn2own.tex
\section{Case Study: Full-chain Exploits}
\label{s:overview:fullchain}
\label{s:pwn2own}

Because modern browsers are heavily compartmentalized with
different security capabilities, browser exploitation often
requires chaining multiple attacks to ultimately
execute malicious action.
Combining all such steps
is usually referred to as \emph{full-chain exploitation}. 
As a representative case study for full-chain browser exploitation,
we analyze a winning attack 
against Safari~\cite{jin:pwn2own2020-safari}
in 2020 Pwn2Own competition, \ie,
the \emph{exploit example} shown in~\autoref{fig:main}.

This attack infiltrates the renderer process, starting from a
JIT compiler optimization error~\cite{CVE-2020-9850}:
The DFG compiler in Safari \js renderer
incorrectly models a side effect of \emph{in} operator
when a special condition regarding \emph{proxy object} is met.
This bug allows the players to construct 
the standard \emph{addrof/fakeobj} primitives,
which yields arbitrary memory read/write 
and ultimately, 
arbitrary code execution.
To construct a valid object using \emph{fakeobj}, 
the players utilize a
publicly known technique~\cite{structure-id-bypass} 
to bypass \emph{object shape authentication}
(\emph{StructureID randomization} in \autoref{s:vuls:jit}).
After faking a JavaScript object, 
they use a known technique~\cite{gigacage-bypass}
to bypass Address Space Isolation 
(\emph{Gigacage} in \autoref{s:vuls:jit})
and get an arbitrary read/write primitive in the renderer process.

Once the renderer process is compromised, 
sandbox escaping is the next step and is more challenging. 
In this attack, the players cleverly stitch
multiple logic/memory errors together to escape the sandbox.
The players first additionally obtain
arbitrary code execution from the \emph{CVMServer}
XPC service (part of the built-in OpenGL framework),
which, though sandboxed, 
has the capability to create \emph{symbolic link},
while the renderer process does not have such capability. 
Also, there is an IPC method in Safari,
\cc{didFailProvisionalLoad()},
that can launch an arbitrary app if 
a symbolic link pointing to the app folder is provided.
By combining them, the players can launch arbitrary apps via Safari.
At this point, the sandbox is successfully breached, 
as they
can execute arbitrary applications outside the renderer sandbox,
similar to a user who launches Safari.

The Pwn2Own example we summarized is specific 
but impactful.
Based on this, 
we describe the full-chain browser exploitation in
a more generic way.
First, to find vulnerabilities in the renderer, 
one can leverage fuzzing
techniques~\cite{park2020fuzzing,han2019codealchemist,holler2012langfuzz}
or manually audit the browser source code.
Discovering an exploitable bug 
would be one of the most challenging steps.
After such a bug is found, 
the next step is to achieve an
arbitrary code execution primitive 
within the renderer process context.
However, taking control over the renderer is only a beginning,
since renderers are confined by the \emph{sandbox} mechanism.
To break out of the sandbox, 
the attacker typically targets
flaws in the browser process,
the OS kernel, or IPC protocols.
Unlike attacking the renderer, 
sandbox escape usually requires
chaining high-level logical exploits against
multiple system components.
Once the sandbox is escaped, 
the attacker can execute an arbitrary
program with an equal security level as the browser,
and full-chain exploit is achieved. \looseness=-1

%% file: discussion.tex
\section{Discussion}
\label{s:discuss}
In this section, we discuss several aspects related to browser security.
There are more discussions in the Appendix.

\subsection{Patch-gapping Problems}
\label{s:patch-gap}

% \PP{Time gap between code patches and releases}
Due to the existence of public repositories and issue trackers,
patches in open source browsers can be published 
before a new release is done and made available to end-users, 
enabling attackers to assess the exploitability of patches.
For example, iOS Safari was exploited due to the 1.5-month patch gap~\cite{jsc-exploits}.
To shrink the gap, Chrome introduced bi-weekly security
updates and reduced the release cycles from six weeks
to four weeks~\cite{chrome-new-release-cycle}.
Firefox holds back pushing security fixes to
the repository before releases~\cite{ff-sec-approval,jsc-exploits}
and recommends not including vulnerability information
in patch commits~\cite{ff-sec-approval}.

\subsection{Homogeneity of Browser Engines}
Many secondary browsers use the same browser engine as the leading browsers 
(\eg, Chrome V8). 
As a result, 
a vulnerability in one browser engine can affect 
other browsers that share it.
Among the 15 most popular browsers~\cite{browser-trends}, 
11 of them are based 
on Chrome's engine (including Microsoft Edge~\cite{edge-chrome}), 
as shown in~\autoref{t:shares}.
When a new version of Chrome is released with bug fixes, it is 
not applied immediately to secondary browsers since there is a time gap 
before secondary browsers integrate them.  

According to the release history of secondary browsers 
there are time gaps before applying released security patches, 
which provides an attack window for the attacker. 
For example, one WebKit bug was exploitable on {PlayStation} firmware 
several months after being reported to the WebKit bug tracker~\cite{ps4-exploit-patch-gapping}. 
This was also an issue in Android,
where apps are shipped with bundled rendering engines
\eg, a UXSS bug 
was reported on {Samsung Internet}~\cite{CVE-2017-17859} around 
one month after being reported to Chromium~\cite{CVE-2017-5124}.
Apple solved this problem in iOS by enforcing
all apps to use WebKit libraries provided by the OS and 
rejecting non-compliant apps in their App Store~\cite{app-store-review-guidelines}. 

Moreover, the use of web browser components 
such as renderers and 
JavaScript engines further extends to 
 applications using frameworks 
such as Electron and Android WebView.
Also, Node.js~\cite{nodejs} and Deno~\cite{deno} 
utilize Google's V8 
engine to enable \js outside the context of browsers 
(\eg, for implementing web servers). 
As a result, 
bugs and exploitations of browser engines 
have a broader impact 
beyond just browsers themselves, expanding the need 
for better defense mechanisms to avoid
catastrophic consequences.

\begin{table}[t]
    \centering
    \footnotesize
    \setlength{\tabcolsep}{3pt}
    \input{tbl/browser-shares}
    \caption{Homogeneity of browser engines. Some browsers ship multiple 
    engines to ensure compatibility of web pages ($\dagger$) or due to 
    specific platform requirements, such as WebKit on iOS($\ddagger$)
    ~\cite{app-store-review-guidelines}.}
    \label{t:shares}
    \vspace{-15pt}
\end{table}

\lesson{The homogeneity of browser engines creates a serious problem;
better patching approaches are needed}{
Due to the homogeneity of browser engines,
browser bugs in one browser engine can affect many
other browsers and applications.
We suggest leading browsers such as Chrome
provide their \js engine as a shared library 
for other apps to use, 
so that it is easier to deploy patches
via over-the-air updates, 
instead of manually integrating patches.
}

\subsection{Bug-finding Tools}
\label{s:bug-finding-tool}
Multiple efforts have been made to develop state-of-the-art tools 
for finding browser engine bugs, which can be mainly divided into two 
categories: fuzzing and static analysis.

\PP{Fuzzing}
Fuzzing is one of the most effective strategies for finding bugs
and has been applied to uncover browser bugs since 2012.
We summarize the papers about browser fuzzers in the past decade in
\autoref{t:tools} (Appendix), 
which includes the bug statistics they have found in Chrome, 
Firefox, Safari, and Edge (based on both ChakraCore and V8), along with 
their key techniques.
These fuzzers choose between two classic modes: mutational fuzzing 
(\eg, Montage~\cite{lee2020montage}) and generational fuzzing 
(\eg, CodeAlchemist~\cite{han2019codealchemist}).
Some fuzzers like LangFuzz~\cite{holler2012langfuzz} and 
DIE~\cite{park2020fuzzing} leverage a mix of both modes coupled with 
coverage feedback. Constructing syntactic and semantic aware inputs 
like DIE~\cite{park2020fuzzing} and LangFuzz~\cite{holler2012langfuzz} 
is useful for generating more crashes.
Some industrial efforts on fuzzing browsers are highly effective on 
finding complex browser bugs. 
For example, ClusterFuzz~\cite{clusterfuzz} 
runs on over 25,000 cores~\cite{clusterfuzz-specs} and found over 
29,000 bugs~\cite{clusterfuzz-github} in Chrome. \looseness=-1

\PP{Static analysis}
Recently, there has been another line of work 
in the fuzzing-dominated 
field of browser bug finding. SYS~\cite{systool}, a first-of-its-kind 
static/symbolic tool for finding bugs in browser code, 
showed that static analysis could be scaled 
for the huge codebases of browsers
by breaking them into small pieces. 
Specifically, SYS uses static checkers to find potential error sites and 
then uses their extensible symbolic execution to analyze those error sites.
Therefore, SYS has highlighted an excellent direction for future works in 
the field of browser bug finding by static analysis.

\lesson{Automated bug-finding tools are great,
but they still need improvement}{
State-of-the-art fuzzers from industry are doing a good job 
of capturing bugs in browsers.
However, despite their good performance, 
such tools still cannot replace 
manual audits, 
which remain the dominant approach 
for finding complex logic bugs.
Thus, more advanced bug-finding techniques are needed 
from academia as well as industry.
}

\subsection{Proactive Mitigations}
% \PP{Proactive mitigations are better.}{
Most existing mitigations are reactive, meaning  
they are implemented 
after an exploit has been found, 
which is not good enough.
It would be ideal if a mitigation could be in place 
before the attack happens
(proactive approach), which can defeat unknown threats. 
% one example
For example, Site Isolation~\cite{reis2019site} 
was originally designed 
to mitigate UXSS attacks using out-of-process \emph{iframes}, 
but it also helped
defeat the Spectre/Meltdown attacks, 
which were found by the researchers 
long after the Site Isolation project started. 
This is a good example of a
proactive approach against unknown threats.
% }

% \PP{Defenders need to create advantages}{
In the game of exploit mitigations, 
defenders can never beat attackers 
because the actions of the defenders are transparent to attackers.
Vendors can change this situation by 
secretly deploying new mitigations,
for example, 
in their sandboxes in a safe browsing infrastructure.
This can also help to detect in-the-wild exploits 
and kill bugs by 
collecting samples that are highly likely to be malicious.
Also, vendors can try more aggressive mitigations 
that are likely to 
affect user experiences in such an environment. 
For example, if \emph{StructureID randomization} 
(\autoref{s:vuls:jit}) was deployed 
in a safe browsing sandbox before 
public announcement,
most JIT exploits involving the 
\cc{fakeobj} primitive would have been detected.
% }

%% file: tbl/browser-shares.tex
\newcolumntype{s}{>{\hsize=0.4\hsize}X}
\newcolumntype{z}{>{\hsize=0.7\hsize}X}
\newcolumntype{m}{>{\hsize=1.1\hsize}X}
\newcolumntype{b}{>{\hsize=1.8\hsize}X}

\begin{tabularx}{\columnwidth}{s b m z}
    \toprule
    Engine & Browser & UI Engine & Server App \\

    \midrule
    Chrome & Chrome, Edge, Opera, UC, Android, 
    360Safe, QQ, Yandex, Whale, 
    Puffin, KaiOS & Electron, \newline Android WebView, \newline Qt WebEngine & Node.js, CouchDB \\

    \midrule
    Safari$\ddagger$ & Chrome, Safari, Edge, Opera, UC, QQ, Whale, Puffin & iOS WebView, \newline Qt WebKit & \\

    \midrule
    Firefox & Firefox & & MongoDB \\

    \midrule
    IE/Edge & Edge Legacy, IE$\dagger$, QQ$\dagger$, Whale$\dagger$ & MSHTML %(Windows WebView) 
    & \\

    \midrule
    Total   & 24 & 6 & 3 \\

    \bottomrule
\end{tabularx}

%% file: conclusion.tex
\section{Conclusion}
\label{s:conclusion}
In this paper,
we present
the first SoK of browser security.
We first provide a unified model
to study the security design of four major browsers,
and present
a 10-year longitudinal study
of browser bugs to study the trends.
Then we introduce prominent bug types,
and present state-of-the-art mitigations.
We also study
a real-world full-chain exploit
from Pwn2Own 2020
in detail.
This paper sheds light
on the area of browser security,
and presents key takeaways
that can enlighten researchers
and browser vendors
on future directions
to improve browser security.

%% file: appendix.tex
% \newpage
\appendix

\section{More Discussions}
\label{append:discuss}
\normalsize
\subsection{Table of Bug-finding Tools}
We summarize the papers about browser 
fuzzers in the past decade in
\autoref{t:tools}.
\begin{table*}[htbp]
    \centering
    \footnotesize
    \input{tbl/tools}
    \caption{Comparison of browser engine fuzzers.}
    \vspace{-13pt}
    \label{t:tools}
\end{table*}
\subsection{Privacy-preserving Browsers}
One of the most concerning privacy leakages in web browsing is
the user's IP address.
As web servers can easily collect and store the IP,
the user's geolocation can be instantly exposed with fine granularity
depending on network circumstances (e.g., NAT).
The Tor browser~\cite{tor-website} addresses this problem 
with the \emph{onion protocol},
which re-routes the user's connection using multiple
random nodes in the Tor network, and each node cannot know the user's identity 
(IP) and the destination at the same time. However, privacy can still be 
breached via \textit{website fingerprinting} techniques by observing 
the encrypted network packet sequences
~\cite{sirinam2018deep,panchenko2016website}.
Another browser, Brave~\cite{brave-website}, prevents websites from tracking 
users by removing all ads and ad trackers contained in websites,
but the user's
browsing history can still be leaked~\cite{tiwari2019alternative,brave-news}.

\subsection{Plugins and Extensions}
\label{s:discuss:plugin}

Plugins and extensions are small software programs that customize the
the browser's functionality by offering a wide variety of features.
Plugins such as Java and Flash 
operate within the context of web pages, 
whereas extensions attach additional features to browsers.
Despite their benefits, plugins are major sources of browser instability
\cite{CVE-2019-6481,CVE-2019-5647}.
Plugins also make sandboxing the renderer process impractical,
as plugins are written by third-parties and browser vendors have no control 
over their access to the operating system.
Also, extensions have special privileges within the browser,
making them appealing targets for attackers
\cite{CVE-2020-6554,CVE-2020-15655,CVE-2020-6809}.

\PP{NPAPI plugins}
NPAPI allows browser vendors to 
develop plugins with a common interface.
When the browser visits a page with an unknown content type, 
it searches for and
loads the available plugin 
to delegate the content processing. 
As a result, attackers can trigger a vulnerability 
by assigning a specific 
content type to a web page 
that fools the browser into loading a specific plugin
 that has a vulnerability.
Attacks on NPAPI plugins had been prevalent over different browsers and 
platforms, especially on Java, Flash, and PDF~\cite{npapi-expkits-ms}.
To mitigate the problem, browsers separated the plugin process from
the browser's main process, namely out-of-process plugin mitigation
\cite{firefox-flash-sandboxing,sabanal2012digging}.
However, plugins could still be used for 
browser exploitation and were accused
of being the reason for performance degradation browser crashes.
As a result, all browsers discontinued support for NPAPI plugins
\cite{ff-flash-roadmap}.

\subsection{Difficulty of Deploying Mitigations}
\label{s:difficulty-mitigations}
It is difficult for browser vendors to deploy mitigations for the 
following reasons:

\PPi{a) Compatibility.} 
Third-party code such as browser plugins 
depend on the browser code to function correctly. 
When introducing browser 
mitigations, it is possible to break third-party code, 
which browser vendors 
have no control over. For example, when trying to 
introduce Win32k lockdown for the
Pepper Plugin API (PPAPI) for Chrome in Windows, 
there was a stability issue 
when applying the patch on Windows 8.1 and below, 
which the Chrome team could 
not track down~\cite{win32k-lockdown}, affecting plugins 
such as Flash, PDFium, and Widevine. As a result, 
PPAPI Win32k lockdown was only 
enabled for Windows 10 and not 
Windows 8/8.1 to avoid stability issues.

\PPi{b) Performance.} 
Adding security mitigations is expensive. 
To mitigate 
security threats, browser vendors sometimes choose to 
trade performance for 
security or vice versa. 
For example, the disabling of \cc{SharedArrayBuffer} (SAB) 
in all modern browsers in early 2018 
as a countermeasure for the Spectre attack, 
as discussed in \autoref{s:mitigation:sc}, 
greatly jeopardizes performance 
because SAB was originally designed to 
achieve lightweight synchronization between workers~\cite{SAB-perf}.

\PPi{c) Security.} 
More code usually means more security vulnerabilities. 
Often, introducing mitigations or patches 
increases the attack 
surfaces. After deploying new patches to browsers, browser vendors
often look for bug reports to 
address the new security issues as soon as possible. 
For instance, Firefox launched a whole new class 
of bug bounties only for 
security vulnerabilities in active 
mitigations~\cite{firefox-bb-extend}.

\PP{Reverted mitigations.}
Some mitigations are deployed temporarily 
to mitigate immediate threats
while better mitigations are being developed.
For example, in the SAB case mentioned above, 
shortly after the introduction of
more robust countermeasures, \ie, Site Isolation and COOP/COEP, 
Chrome and Firefox re-enabled the use of SAB
~\cite{chrome-sab-reenable,ff-sab-reenable}. 
Despite all the efforts to ensure that 
the mitigations are safe, performant, 
and compatible, sometimes mitigations have to be 
rolled back due to some severe
consequences they introduce. For example, 
in \autoref{t:mitigation}, XSS Auditor~\cite{xss-auditor}, 
an inbuilt XSS filter
for Chrome, suffered from many security side-effects, 
which led to its 
retirement in 2019~\cite{xss-auditor-retired}.

%% file: tbl/tools.tex
\newcommand{\cmark}{\ding{51}}
\newcommand{\xmark}{\ding{55}}

\raggedleft\textbf{G:} Generational, \textbf{M: } Mutational, \textbf{SM: } Semantic Aware, \textbf{SN: } Syntactic Aware, \textbf{Cov: } Coverage Feedback, \\
\textbf{OS: } Open Source, \textbf{C: }Chrome, \textbf{FF: }Firefox, \textbf{S: }Safari, \textbf{E$\dagger$: }Edge based on V8, \textbf{E$\ddagger$: } Edge based on ChakraCore

\resizebox{\textwidth}{!}{
\begin{tabular}{l c c c c c c @{\hspace{0.5\tabcolsep}} c @{\hspace{0.5\tabcolsep}} c @{\hspace{0.5\tabcolsep}} c @{\hspace{0.5\tabcolsep}} c @{\hspace{0.5\tabcolsep}} c @{\hspace{1.5\tabcolsep}} l }
        \toprule
        Fuzzer & Year & C+E$\dagger$ & FF & S & E$\ddagger$ & G & M & SM & SN & Cov & OS & Key Techniques\\
        %\cmidrule(lr){3-6} \cmidrule(lr){7-8}

        \midrule
        SoFi~\cite{sofi} & 2021 & 1 & 5 & 1 & 18 & \xmark & \cmark & \cmark & \cmark & \cmark & \xmark & Uses fine-grained program analysis and repair strategy to generate semantically valid inputs \\

        Token-Level Fuzzing~\cite{sallstokenlevel} & 2021 & 16 & 3 & 4 & 6 & \xmark & \cmark & \xmark & \xmark & \cmark & \xmark & Applies mutation at \textit{token-level} by changing or replacing~entire words \\
         
        Favocado~\cite{dinh2021favocado} & 2021 & 8 & NA & 5 & NA & \xmark & \cmark & \cmark & \cmark & \xmark & \cmark & Generates test cases based on semantic information, and tracking states mutation\\
         
        DIE~\cite{park2020fuzzing} & 2020 & 4 & NA & 16 & 28 & \cmark & \cmark & \cmark & \cmark & \cmark & \cmark & Preserves beneficial properties and conditions called \textit{aspects} across mutation \\

        FREEDOM~\cite{xu2020freedom} & 2020 & 4 & 5 & 13 & NA & \cmark & \cmark & \cmark & \cmark & \cmark & \cmark & Uses customized IR (FD-IR) to describe HTML documents and to define mutation\\ 

        Montage~\cite{lee2020montage} & 2020 & 1 & 0 & 2 & 34 & \xmark & \cmark & \cmark & \cmark & \xmark & \cmark & Transforms JS ASTs into sequences to train Neural Network Language Models (NNLMs)\\
        
        Nautilus~\cite{aschermann2019nautilus} & 2019 & NA & NA & NA & 2 & \xmark & \cmark & \xmark & \cmark & \cmark & \cmark & Combines grammar-based input generation with coverage feedback\\
        
        Deity~\cite{lin2019deity} & 2019 & NA & NA & 1 & 1 & \cmark & \cmark & \xmark & \cmark & \cmark & \cmark & Generates syntactic JS code using previously known bugs and PoCs\\
        
        Superion~\cite{wang2019superion} & 2019 & NA & NA & 16 & 3 & \xmark & \cmark & \xmark & \cmark & \cmark & \cmark & Employs grammar-aware test input trimming with tree and dictionary-based mutation\\
        
        CodeAlchemist~\cite{han2019codealchemist} & 2019 & 2 & NA & 10 & 7 & \cmark & \xmark & \cmark & \cmark & \xmark & \cmark & Tags code bricks with constraints defining when to combine with other code bricks\\
        
        LangFuzz~\cite{holler2012langfuzz} & 2012 & 11 & 20 & NA & NA & \cmark & \cmark & \cmark & \cmark & \xmark & \xmark & Generates grammar-aware test inputs, and leverages previously known faulty programs\\

        \bottomrule
        Total &  & 42 & 23 & 54 & 81 & &&&&&& \\
        \bottomrule

\end{tabular}}